\documentclass[10pt]{IEEEtran}

\usepackage{amsmath}
\usepackage{amsthm}
\usepackage{amsfonts}
\usepackage{amssymb}
\usepackage[australian]{babel}
\usepackage{fontenc}
\usepackage{graphicx}
\usepackage{yhmath}
\usepackage{epstopdf}
\usepackage{color}
\usepackage{subfigure}
\usepackage{algorithm}

\newtheorem{lem}{Lemma}
\newtheorem{prop}{Proposition}
\newtheorem{rem}{Remark}

\title{Higher order mobile coverage control with application to localization} % Title, preferably not more
\author{Bomin Jiang, Zhiyong Sun, Brian D. O. Anderson and Christian Lageman
\thanks{This work  is supported by the Australian Research Council's Discovery Projects DP-110100538,  DP-130103610 and DP-160104500,  by NICTA-Data61, and a Germany-Australia DAAD-Go8 Collaborative Award. Z. Sun is also supported by the Prime Minister's Australia Asia Incoming Endeavour Postgraduate Award. }
\thanks{Bomin Jiang is with Laboratory for Information and Decision Systems, and Department of Mechanical Engineering, Massachusetts Institute of Technology. (e-mail: bominj@mit.edu).  Zhiyong Sun and Brian D. O. Anderson are with Research School of Engineering, The Australian National University. Canberra, Australia. (e-mail: \{zhiyong.sun, brian.anderson\}@anu.edu.au). Christian Lageman is with Institute of Mathematics, University of W{\"u}rzburg, W{\"u}rzburg, Germany. (e-mail: christian.lageman@mathematik.uni-wuerzburg.de).}
}

\begin{document}

\maketitle

%\begin{keyword}                           % Five to ten keywords,
%Multi-agent systems; Distributed control; Sensor networks; modeling for control optimization; mobile robots; Localization.             % chosen from the IFAC
%\end{keyword}                             % keyword list or with the
                                          % help of the Automatica
                                          % keyword wizard

\begin{abstract}
Most current results on coverage control using mobile sensors require that one partitioned cell is associated with precisely one sensor. In this paper, we consider a class of coverage control problems involving higher order Voronoi partitions, motivated by applications where more than one sensor is required to monitor and cover one cell. Such applications are frequent in scenarios requiring the sensors to localize targets.  We introduce a framework depending on a coverage performance function incorporating higher order Voronoi cells and then design a gradient-based controller which allows the multi-sensor system to achieve a local equilibrium in a distributed manner. The convergence properties are studied and related to Lloyd algorithm. We study also the extension to coverage of a discrete set of points. In addition, we provide a number of real world scenarios where our framework can be applied. Simulation results are also provided to show the controller performance.
\end{abstract}

\section{Introduction}

%In recent years, the performance of multi-agent systems (MASs) in various tasks, e.g. mobile coverage control, consensus \cite{ren2007information,olfati2007consensus}, formation shape control \cite{ren2010distributed, anderson2008rigid,6760298}, cooperative geolocalization \cite{wymeersch2009cooperative}, etc. has been studied with increasing intensity.
Consider a closed region, for convenience convex and polygonal. Suppose that across this region, individuals are distributed, in general nonuniformly. Computational geometry \cite{de2000computational} then deals with questions such as: how can $n$ supermarkets be located so that the sum of the trip distances for every individual to his or her nearest supermarket is minimized? (Such a problem is termed a `post office' problem in \cite{de2000computational}). See also \cite{clarkson1985probabilistic}. In recent years, mobile versions of this sort of problem have been considered, where there is a set of mobile agents that must cooperate to find optimum locations to extremize an index of the supermarket location type. Indeed, a fundamental problem is the optimal positioning of agents (where \emph{agents} may refer to mobile sensors or autonomous vehicles) to cover an area in a way that some predefined coverage performance function can be optimized. This performance function  can be related to the quality of service of a mobile sensing network, or the cumulative probability of certain events detected by sensors in the area of interest.

Now consider the following variant on the supermarket problem \cite{martin2009review}(a variant that was actually embedded in legislation, now discarded, of the Australian Capital Territory): where should the $n$ supermarkets be located in order that the sum over all individuals of the sum of the distances to the two closest supermarkets is minimized? (The social motivation was to ensure that no one supermarket had an approximation to a locational monopoly). This is an example of a higher order coverage problem. Naturally too, one can conceive of a version in which there are mobile agents, rather than supermarkets, and the agents must cooperate to reach their optimal positions.

Problems of this nature have in fact been identified to us by the Australian Government  Defence Science and Technology Group as being of significant relevance. There is a fundamental problem of optimal positioning of agents (e.g. mobile sensors on the ground, unmanned airborne vehicles, etc.) to cover an area which is important in a range of applications problems, including maximizing the quality of service of a mobile sensing network \cite{cortes2004coverage}, and maximizing the probability of  detecting an intruder, and indeed  localizing an intruder. This task frequently involves more than one agent since many localization approaches rely for example on such notions as triangulation using simultaneous bearing measurements of the target from different locations, or exploiting time difference of arrival of signals emitted by or reflected from the target.
In order to achieve the positioning optimization goal, it is usually required that the control algorithms for different agents are executed in a decentralized manner. That is to say, agents are autonomous, being capable of making decisions based on their own information, some of which could be obtained from neighboring (normally geographically close) agents.
%Coverage is one of the most fundamental tasks for mobile sensor networks,
This type of mobile coverage control problem has been studied extensively in the literature since \cite{cortes2004coverage} for the case where only a single agent is required to cover each point in the area of interest. Our goal in this paper is to deal with the much less well studied problem of multiple coverage.
%The  coverage control task requires a large number of mobile sensors to coordinate their actions and to monitor an environment. In coverage control problems, a group of mobile robots (sensors) are to be positioned in a bounded and convex environment.
%There are many papers on various extensions on this problem including obstacle avoidance \cite{lindhe2005flocking}, non-convex area coverage control \cite{breitenmoser2010voronoi}, coverage control with time-varying density functions \cite{lee2013controlled}.

The typical current mobile coverage control (involving single agent coverage of a given cell) problem is usually solved by using the geometric tool of  Voronoi partitions \cite{de2000computational}. When using this tool, each agent is thought of as monitoring a convex area whose shape is determined in part by information from its neighbouring agents. Furthermore, each agent moves according to some gradient-based control law to achieve optimization of the coverage performance criterion. There is also some literature treating the $k$-coverage problems, e.g.,  \cite{li2010distributed,so2005solving}. In the paper \cite{li2010distributed} it converts the $k$-coverage problems to a 1-coverage problem, then uses a distributed $k$-coverage self-location estimation (DSLE) scheme based on the Voronoi diagram to implement coverage. Our paper, on the contrary,  tackles  the $k$-coverage problem directly using the concept of higher-order Voronoi diagrams. In another direction, focusing on computational complexity analysis, the paper \cite{so2005solving} considers the problem of verifying $k$-coverage of a given region with fixed-position sensors. However, mobile coverage control, which is the emphasis of our paper, is not studied in \cite{so2005solving}.

In considering the generalization of the classical coverage problem to consider monitoring of each location by more than one agent,  the starting
idea is to extend `one agent being responsible for one cell' to `two or more agents being responsible for one single cell'. This generalization is motivated by
many real-world applications. As alluded to above, in bearing-only sensing, each sensor determines the direction but not the range of an object of interest; determination of the object's position thus requires detection by at least two sensors. To avoid potential problems due to collinearity of a target object and two sensors, coverage may in fact require three sensors for each cell. Again, in geolocation using time-difference of arrival (TDOA), data received by two sensors suffices to locate an object on one branch of a hyperbola. To locate the object, at least one further  independent TDOA measurement (thus requiring a third sensor) is needed; this provides a second hyperbola branch on which the target object is located. In later discussions in the paper, yet another example again is given, of a bistatic radar\cite{4201772}.
The usual current framework for considering coverage control with a single agent is not immediately suitable for the above applications.

In this paper, we will use the concept of a \textit{higher order Voronoi partition} as the main tool for tackling the problem. This concept is not a recent idea. The book \cite{okabe2009spatial} mentioned the concept of a higher order Voronoi partition and the paper \cite{6258029} considered a wireless sensor optimization problem requiring use of the concept. Nevertheless, \textit{mobile} coverage control using higher-order Voronoi partitions is novel to the best of our knowledge.
One of the techniques for solving one class of conventional (first order) Voronoi partition problems (those which specifically work with the square of Euclidean distance) is known as Lloyd's algorithm. For such first order problems, the optimal solution results in the so-called \textit{generator} of each Voronoi set (the term is explained below) being located at the centroid of that set. The solution algorithm, which is not a basis for determining smooth motions of mobile sensors but simply an algorithm for determining optimal positions which they should attain, proceeds iteratively. We study in this paper a higher order version of this algorithm.

Some of the ideas of this paper were introduced in a much more limited predecessor work, \cite{JSA2015}. However new material going well beyond the predecessor work is included here.  In particular, we include discussions on higher order Lloyd algorithms as just mentioned, as well as minimum sensing radius problems, collision avoidance, convergence issues, and discrete set coverage.

The rest of this paper is organized as follows. Section \ref{Sec:background} reviews the problem settings and current results on the coverage control problem, and further presents a brief introduction to order $k$  Voronoi partitions. However,  as compared with \cite{JSA2015}, there is now included material on coverage problems aimed at minimizing the sensing radius in a sensing network covering a region.   Section  \ref{Sec:order2} discusses mainly the order 2 coverage control problem, the controller design,  and the stability analysis of the coverage sensor system.  While some of this material appeared in \cite{JSA2015}, the last four subsections covering the Lloyd algorithm, minimum sensing radius coverage, collision avoidance and convergence issues are new. There is also new material in the earlier subsections  on the computation of gradients and Hessians.  Section \ref{Sec:higher_order} deals with coverage problems for finite sets, with no corresponding material having been included in the predecessor paper. This generalizes what is generally known as $k$-means clustering. Simulation results are provided in Section \ref{Sec:simulation} to show the coverage properties of various controllers. In Section \ref{Sec:application}, we show some potential applications of the high order coverage framework.  Finally, Section \ref{Sec:conclusions} concludes this paper.

\section{Background literature} \label{Sec:background}
This section gives a review of current order-1 mobile coverage control, followed by the tools of higher order Voronoi diagrams and the Lloyd algorithm. Furthermore, a variant on using the earlier performance index is also considered. 

\subsection{Order 1 Voronoi partition and coverage control}  \label{SubSec:order1}
\label{Background}
Suppose there is a 2-D convex area $Q$ to be covered by $n$ mobile sensors in this area.
A point in $Q$ is denoted as $q$ and sensor $i$'s position is denoted by $p_i \in \mathbb{R}^2$. A coverage performance function $\mathcal{H}(p_1, p_2,\cdots,p_n)$ is defined as follows
\begin{equation}
\label{single performance define}
\mathcal{H}(p_1, p_2,\cdots,p_n)=\int_Q \min_{i\in \{1,\cdots,n\}} f(\|q,p_i\|) \phi(q)dq
\end{equation}
where $\phi$ is a distribution density function known to all sensors which is assumed to be $C^2$ (i.e. two times continuously differentiable), $\|q,p_i\|$ denotes the Euclidean distance between $q$ and $p_i$, and
the function $f(\|q,p_i\|)$ describes the measurement cost or quantitative measurement assessment of how poor the sensing performance is  at a point $q$ by a sensor at $p_i$. Thus larger values correspond to poorer sensing. We also suppose that $f$ should be $C^2$ and  monotonically increasing. The coverage control aims to minimize the above performance function and to find and achieve the corresponding optimal positions of (mobile) agents.

The minimum inside the integral of the performance function \eqref{single performance define} induces a partition of $Q$ into non-overlapping cells. These cells   are called the {\it{Voronoi partition}} $\{V_1,\cdots,V_n\}$ of $Q$ {\it{generated}} by the points $p_1, p_2,\cdots,p_n$ defined as
\begin{equation}
V_i=\{q\in Q | ~\| q,p_i\| \leq \| q,p_j\|,\forall j\neq i\}
\end{equation}
The individual Voronoi cells are all convex, and polytopic when $Q$ is polytopic. The boundary of two adjacent cells is the perpendicular bisector of the line segment joining the generators of the cells. For a given set of $p_1, p_2,\cdots,p_n$, if $V_i$ and $V_j$ are adjacent (i.e. two cells which have a boundary comprising an interval of nonzero length), then agents  $i$ and  $j$ are defined as {\it{neighbors}}.
%{\color{red} Zhiyong has raised the following point and it is being chased down.  There are nongeneric situations where four boundary lines meet at the same point. There are then generators for which the (closures of the ) associated Voronoi regions do not share a common line segment but they do share a common point. It is not clear whether such generators should or should not be determed neighbors [Bomin: if the purpose that we define neighbours is for the distributed computation of gradient/Lloyd algorithm, then two agents sharing a common point should not be considered as neighbours if they are otherwise not sharing a line. Because the computation of boundaries does not requires any information exchange between them.] [Zhiyong: I have checked several papers (and bookds) on definition of Voronoi neighbors and found Bomin's interpretation is correct. In particular in Cortes' 2004 and 2005 SIAM paper, it's mentioned "sites are Voronoi-neighbors if they share an edge, not just a vertex.". Please delete the red part if this point is confirmed and agreed. ]  }
The {\it{neighbor set}} of agent $i$ is denoted as $\mathcal{N}_i$.
By using the Voronoi partition, the performance function \eqref{single performance define} can be transformed as
\begin{equation}
\int_Q \min_{i} f(\|q,p_i\|) \phi(q)dq=\sum_{i=1}^n \int_{V_i} f(\|q,p_i\|) \phi(q)dq
\end{equation}
The controller proposed in e.g. \cite{cortes2004coverage} is a gradient-based controller minimizing the performance function, i.e. a control law is provided to smoothly move the generators $p_i$ to minimize the performance function. An optimal coverage performance can be obtained by moving mobile sensor positions in accordance with the gradient-based law. The optimum may be local, not global. Below we explore in more detail aspects of the equilibria.  In the rest of this paper, we term this an \emph{order 1 coverage control} problem. Order 1 coverage control problems stand in contrast to order $k$ coverage control problems, which we now introduce.

\subsection{Order $k$ Voronoi partition}
\label{order k voronoi partition}
As noted already, most literature on coverage control, such as the work reviewed in Section \ref{SubSec:order1}, assumes that each sensor is responsible for sensing or monitoring its own region,
%As mentioned above, we are going to consider multiple mobile sensors in an area for the first time. The starting idea is to extend `one robot in each partition' to  `two or multiple robots in each partition'. In the literature there is a relevant concept called `higher-order Voronoi Partition'.  This is not a recent idea, and an early paper even appeared in the 1970s \cite{shamos1975closest} in which the connection between order 2 Voronoi diagram and closest-point pair problems was discussed. The above problem is an order one coverage control problem in the sense that each subset $V_i$ only corresponds to one sensor.
and we have an \emph{order 1 coverage control} problem.
Now we  generalize the problem to a higher order coverage problem such that each cell is defined by two or more sensors.
%As a matter of fact, although the order $k$ coverage control problem (when $k\geq 2$) is novel, the order $k$ Voronoi partition problem is not.
The general literature of order $k$ Voronoi partitions includes \cite{aurenhammer1991voronoi,boissonnat1993semidynamic}. Mature algorithms are also available to compute the order $k$ Voronoi partition of a given area; see the survey paper \cite{aurenhammer1991voronoi} or \cite{agarwal1998constructing}. Note that most algorithms reported in these earlier papers are centralized ones, but a distributed algorithm has appeared in some recent papers (see e.g. \cite{6258029}).

The definition of an order $k$ Voronoi partition of a convex area $Q$ is given below following \cite{agarwal1998constructing}. Let $\mathcal{S}$ be a finite set of sensors' positions in $Q$. Suppose further that $\mathcal{T}$ is a subset of $\mathcal{S}$ and there are $k$ elements in $\mathcal{T}$.  The generalized Voronoi partition is defined by the collection of subsets of $Q$:
\begin{equation}\label{define order k}
V_{\mathcal{T}}=\{q |\forall v\in \mathcal{T}, \forall w\in \mathcal{S} \backslash \mathcal{T}, \|(q,v)\| \leq \|(q,w)\|,|\mathcal{T}| = k\}
\end{equation}
where $\mathcal{S}\backslash\mathcal{T}$ denotes the relative complement of $\mathcal{T}$ with respect to $\mathcal{S}$. For each point $q$ in $V_{\mathcal{T}}$, $q$ is not further to any sensor in $\mathcal{T}$ than to any sensor not in $\mathcal{T}$. The set $\mathcal T$ is the generator or generating set of $V_{\mathcal T}$.

In this paper we largely focus on the order 2 coverage problem.
%In future work, we can generalize it to order $k$ cases.
As an example, Figure 1(a) shows an order 1 Voronoi partition and Figure 1(b) shows a corresponding order 2 Voronoi partition with the same sensor positions as  Figure 1(a). There are some similar properties and some dissimilar properties in higher-order partitions in comparison to order 1 Voronoi partitions, and we list some, beginning with similarities:
\begin{enumerate}
\item No two cells of $V_{\mathcal{T}}$ overlap, except on their boundaries.
\item The union of the cells, which are all closed, covers the convex region $Q$.
\item  All cells are convex.
\item Boundaries of neighboring cells in an order $k$ partition are line segments each of which is collinear with the perpendicular bisector of a line joining two agents which are neighbours in the order $k$ partition.

\item Not every $k$-combination of sensors necessarily defines a cell in the partition, i.e. for some $\mathcal T$, the subset $V_{\mathcal T}$ may be empty.
\end{enumerate}

\begin{figure}
\begin{center}
\subfigure[Order 1 Voronoi partition]{
\includegraphics[width=0.46\columnwidth]{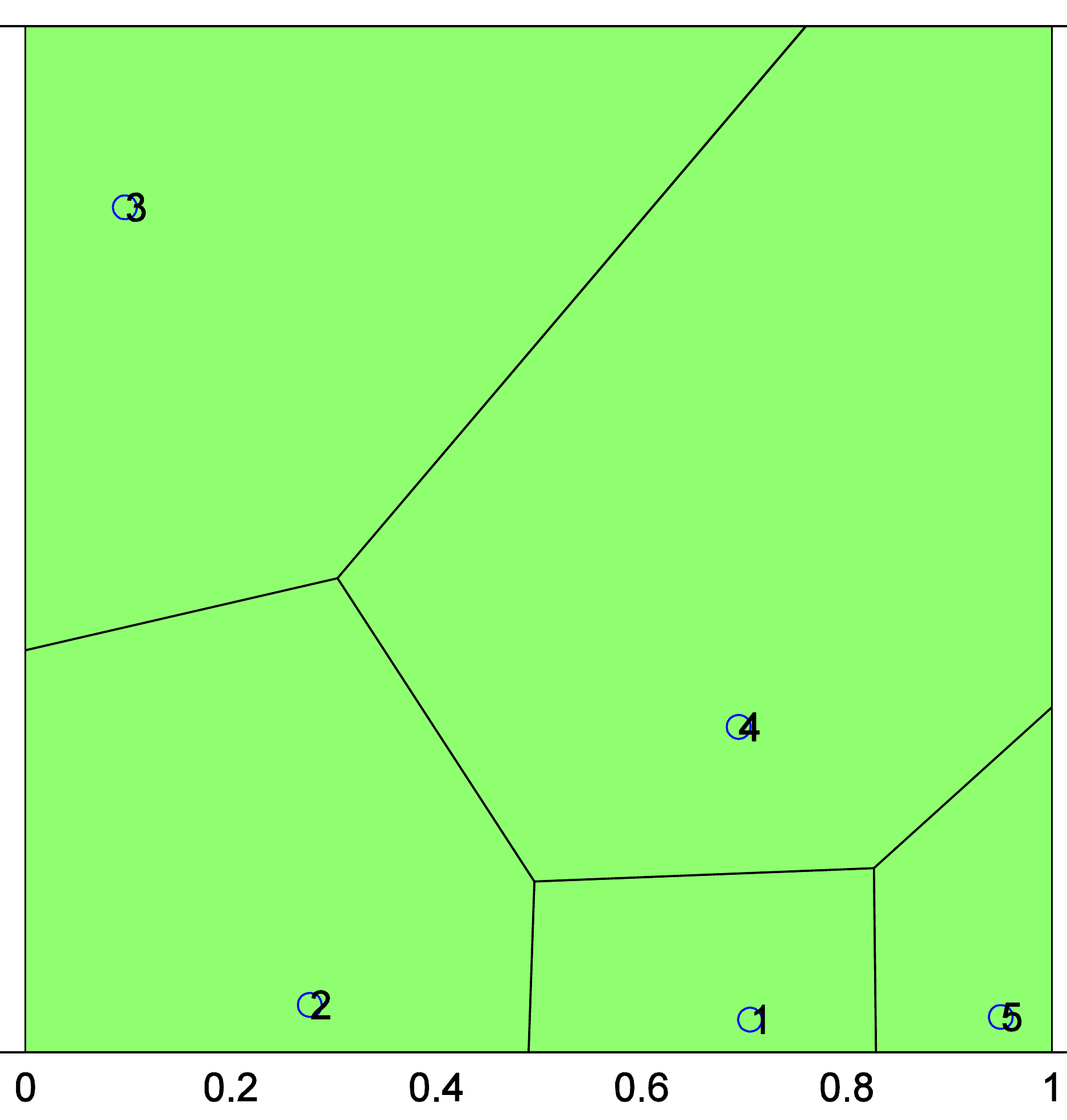}
}
\subfigure[order 2 Voronoi partition]{
\includegraphics[width=0.46\columnwidth]{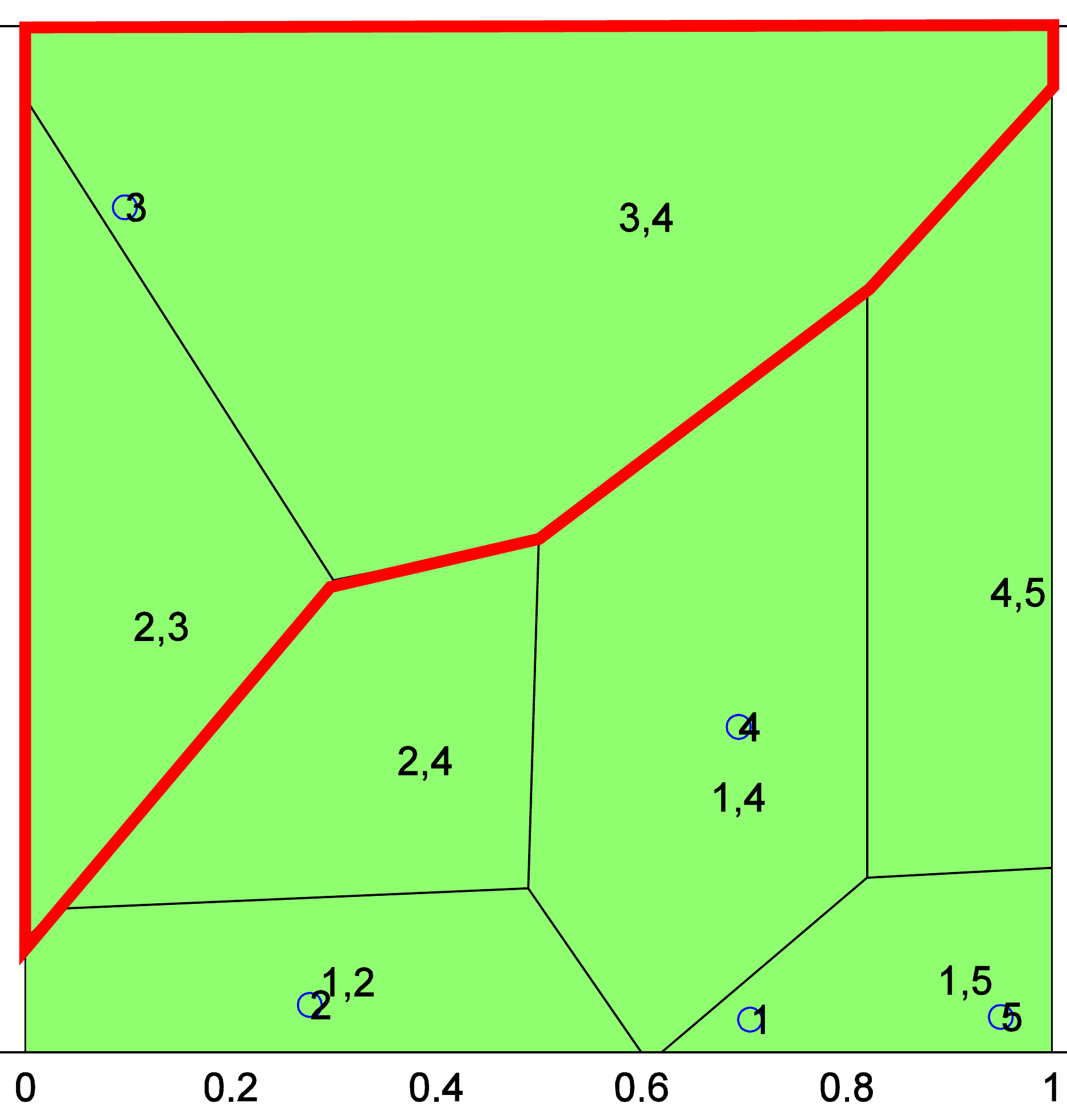}
\label{fig2}
}
\end{center}
\caption{Examples of order 1 and 2 Voronoi partitions with the same sensor positions. The partitions are not centroidal and the generators are depicted, rather than the centroids of each cell}
\label{a2}
\end{figure}

For each $p_i\in \mathcal{S}$, there are some $\mathcal{T}$ that contain $p_i$. We put all these $\mathcal{T}$ into a set $\mathcal{P}_i$ so that $\mathcal{P}_i=\{\mathcal{T}| \mathcal{T} \subset \mathcal{S},p_i\in\mathcal{T}\}$ and then the union of cells with the sensor $p_i$ common to all cells is $W_i=\cup_{\mathcal{T}\in\mathcal{P}_i} V_{\mathcal{T}}$. It is noticeable that when we put these $\mathcal{T}$ together, we will not obtain the same cell containing $p_i$ in the order 1 partition. In fact, there holds $p_i\in V_i\subset W_i$. In addition, according to the definition of higher order Voronoi partition, $V_i$ and $V_{\mathcal{T}}$ are both always convex but $W_i$ may not be convex. An illustration  is given in Fig. \ref{fig2}, where the red contour comprises $W_3$. Other properties of higher order Voronoi partitions can be found in \cite{chazelle1987improved} and \cite{lee1982k}.

\subsection{Centroidal Voronoi partitions and the Lloyd algorithm}
For order 1 Voronoi partitions, given the choice $f(q,p_i)=||q-p_i||^2$, it is easily checked that for a fixed partition of the set $Q$, the optimal $p_i$, i.e. those minimizing the performance function, are located at the centroids of the different cells. Conversely, if the generator positions are fixed, the optimum shape cells are defined by a Voronoi partition. Unsurprisingly, when optimization occurs both over the generator positions and the cell boundaries, the optimum has the generators coinciding with the cell centroids, and is known as a centroidal Voronoi partition, Such partitions have been heavily studied, see e.g.  \cite{du1999centroidal,du2010,liu2009}

Not every centroidal Voronoi partition is however optimum, either globally or locally. Multiple minima and saddle points can arise. As we shall see, the idea of centroidal Voronoi partitions will also be useful in looking at order $k$ partitions.

A key feature of the choice of $f(q,p_i)=||q-p_i||^2$ for which optimality arises with the centroidal Voronoi partition is the elegant form of a gradient descent algorithm for finding stationary points of the performance function. It is simply

\begin{equation}\label{eq:orderonemotion}
\dot p_i=\alpha M_{V_i}(c_i-p_i)
\end{equation}
where $\alpha$ is a positive constant,  $M_{V_i}$ is the mass of the $i$-th Voronoi cell,  and $c_i$ is the centroid of the Voronoi cell whose generator is $p_i$ (for more details and definitions, see e.g. \cite{du1999centroidal}). Note that $c_i$ is not a constant; when $p_i$ changes, the associated cell changes and then its centroid may change.

Partly because it is a comparatively straightforward task to find the centroid of a convex polytope, a recursive algorithm known as Lloyd's algorithm originally developed in another context \cite{lloyd1982} provides another approach to reaching an optimum partition. In this algorithm, one alternates between finding the centroids of a Voronoi given partition, and finding the Voronoi partition corresponding to a given set of generators, see e.g  Section 5.2 of \cite{du1999centroidal}. It is shown in \cite{cortes2004coverage} that one step of the Lloyd algorithm causes the performance index to either decrease, or remain the same; this paper even uses the term {\it{continuous time Lloyd descent}} to describe a gradient descent algorithm computed for the performance index.

These ideas will all be reflected below in the study of higher order coverage control.

\subsection{Minimizing sensing radius in a coverage problem}
There is a variant on using the earlier performance index which is relevant to consider. It gives rises to an iterative algorithm like the Lloyd algorithm with associated Voronoi partitions, and it has arisen in the context of wireless coverage problems, see e.g. \cite{6258029,So2005}.  We introduce the order 1 version now, in preparation for treating the higher order version later.

Suppose first that we start with a set of sensor positions, and seek to find the optimum partition such that the maximum over all cells of the maximum distance of the generator to any point in its cell is minimized. A Voronoi partition based on taking the sensor positions as generators is optimum (but other non-Voronoi optima may exist too).

Secondly,  suppose that we have constructed an order 1 Voronoi partition for a fixed set of generators, and then we ask the question: for this partition, where are the sensor positions inside the subsets of the partition that would allow sensors of the minimum sensing radius over all possible sensor positions (all sensors being assumed to have the same sensing radius and to cover the region)?  This is equivalent to finding the Chebyshev centers of the cells, i.e. the centers of the minimum radius circumcircles \footnote{There are two different definitions of  Chebyshev center, i.e. the center of the minimal-radius ball \emph{enclosing} the entire set, or alternatively (and non-equivalently) the center of largest \emph{inscribed} ball of the convex set. We refer the readers to \cite{cortes2005coordination} for discussions and illustrations of these two different centers. In this paper we consider the former definition.}, and with convex polygonal cells, these are uniquely defined and readily found, see e.g., \cite{wu2013,eldar2008}.

One can go further and ask: what is the optimum partition for which the sensors can use the minimum sensing radius?
By analogy with the centroidal Voronoi partition concept associated with the weighting function $f(q,p_i)=||q-p_i||^2$, there is a simple answer: it will be a Voronoi partition in which the generators coincide with the Chebyshev centers of the cells. And by analogy with the Lloyd algorithm, an iteration based on alternating between finding the Voronoi partition corresponding to a set of generators and finding the Chebyshev centers for the cells in a partition can be used to converge to an optimum. It is easily shown that one step of the algorithm either leaves the required sensing radius unchanged, or reduces it; hence the algorithm converges.   By analogy with the form of the continuous time descent algorithm for $f(q)=||q-p_i||^2$, we suggest use of the algorithm
\begin{equation}\label{eq:ctscheb}
\dot p_i=\alpha(c_i^{cheb}-p_i)
\end{equation}
where $c_i^{cheb}$ is the Chebyshev center of the $i$-th subset of the Voronoi partition, and $\alpha$ is a positive constant. Such
a continuous-time algorithm \eqref{eq:ctscheb} has been established and discussed in \cite{cortes2005coordination} (see Eq. (5.8) of that paper). Note that \eqref{eq:ctscheb} is not necessarily a
gradient-descent system associated with some cost performance function, while a gradient-descent system involving non-smooth analysis for reaching the Chebyshev circumcenter of a convex region is discussed in detail in \cite{cortes2005coordination}.  Also note that the right-hand side of  \eqref{eq:ctscheb} is continuous, because the Chebyshev circumcenter of a polygon depends continuously on the locations of its vertices,  and the location of the vertices of the Voronoi partition depends continuously on the location of the generators. This property is established for order 1 Voronoi partitions in  \cite{okabe2009spatial,cortes2005coordination}.
% We will later show that the same continuity statement also holds for the higher order algorithm.

%We are not asserting that this is a steepest descent algorithm corresponding to a particular index reflecting the minimum sensing radius, but simply that this is a well-motivated algorithm.
A higher order version, suggesting an algorithm analogous to a Lloyd algorithm, is proposed in \cite{6258029}.
Of course, analogously to the usual Lloyd algorithm, the quantities $c_i^{cheb}$ depend on the $p_j$. The behavior of this algorithm for the $k$-th order case is considered later in the paper.
%{\color{red} Questions for Bomin. DO you know other  references for the first order version of the problem statement, and the associated iterative Lloyd-like algorithm? As you know, we mention the higher order version later in the paper. Also, is there treatment in the literature of a continuous time gradient-like algorithm for this problem, as opposed to the iterative algorithm? [Bomin: the gradient is the vector from current point to the centroid. Therefore the iterative method may be equivalent to the gradient one. ] Also, feel free to adjust any of my description above, but preferably use blue or red for the new stuff}

\section{Order 2 coverage control}  \label{Sec:order2}
\label{order 2 section}
In this section, we   discuss order 2 coverage control, so throughout this section we assume that $k=2$. However we note that the analysis of controller design and the convergence property can also be extended to more general order $k$ coverage control, and in the next section we will provide more detailed discussions on the generalization. To clarify notations for this section, define the set $C=\{i,j|~i,j\in\{1,\cdots,n\},i < j\}$. We also note that the set designation of a particular $\mathcal{T}$ as $\mathcal{T}_{ij}$ means points $p_i,p_j \in \mathcal{T}_{ij}$ and $(i,j)\in C$.

\subsection{Performance function and its relationship with generalized Voronoi partition}
\label{Performance function and its relationship with generalized Voronoi partition}
The generalized sensing performance function $\mathcal{H}(p_1, p_2, \cdots, p_n)$ is constructed  as follows
\begin{equation}
\label{double cover}
\mathcal{H}(p_1, p_2,\cdots,p_n)=\int_Q \min_{(i,j)\in C} f(\|q,p_i\|,\|q,p_j\|) \phi(q) dq
\end{equation}

We are trying to find optimal positions of the sensors that can minimize the above performance function. Similarly to an order 1 Voronoi coverage problem, the function $f(\cdot, \cdot)$  indicates the measurement quality of a point $q$ but now reflecting a pair of agents.
Therefore, for a set of fixed sensor positions, we might measure the quality of sensing associated with those positions by $\int_Q \min_{(i,j)\in C} (\|q,p_i\|+\|q,p_j\|) \phi(q) dq$.  In fact, there is a broad set of $f(\cdot, \cdot)$ for which it makes sense to formulate such a measure. We shall in fact impose certain properties on $f(\cdot, \cdot)$ analogous to the order 1 mobile coverage control case, where $f(\cdot)$ being monotonically increasing and differentiable is a basic requirement. In the order 2 case, besides requiring that the function $f(\cdot, \cdot)$ should be differentiable, it should also have the following properties

\begin{enumerate}
  \item $\frac{\partial}{\partial \|q,p_i\|}f(\|q,p_i\|,\|q,p_j\|)\geq 0$ \\
  \item $\frac{\partial}{\partial \|q,p_j\|}f(\|q,p_i\|,\|q,p_j\|)\geq 0$, and \\
  \item $f(\|q,p_i\|,\|q,p_j\|)=f(\|q,p_j\|,\|q,p_i\|)$
\end{enumerate}
The first two properties correspond to the monotonically increasing property in the order 1 case. We note that the higher order Voronoi partition is defined as
\eqref{define order k}, where the order of the elements in $\mathcal{T}$ should not affect the actual partition. As a result,  the order of independent variables should not affect the value of $f(\cdot, \cdot)$, either. The third requirement in the above condition ensures this property.

Based on the above performance function \eqref{double cover} and the distance function, we can obtain the following two lemmas.

\begin{lem}
\label{lemma1} With the definitions of $Q$, $C$, $f(\cdot, \cdot)$ $\mathcal{T}_{ij}$, $\mathcal{S}$ and $V_{\mathcal{T}_{ij}}$ stated above, for all $q$ in the set $V_{\mathcal{T}_{ij}}$ and  $(k,l)\neq(i,j)$, there holds \begin{equation} f(\|q,p_i\|,\|q,p_j\|) \leq f(\|q,p_k\|,\|q,p_l\|)\end{equation}
\end{lem}

\parskip= 6pt
\begin{proof}
According to the definition of $V_{\mathcal{T}_{ij}}$, we know each $\|q,p_i\|$ and $\|q,p_j\|$ is less than or equal to both $\|q,p_k\|$ and $\|q,p_l\|$. Because $\frac{\partial}{\partial \|q,p_i\|}f(\|q,p_i\|,\|q,p_j\|)\geq 0$, there holds $f(\|q,p_i\|,\|q,p_j\|)\leq f(\|q,p_k\|,\|q,p_j\|)$.

Further because $\frac{\partial}{\partial \|q,p_j\|}f(\|q,p_k\|,\|q,p_j\|)\geq 0$, there also holds $f(\|q,p_k\|,\|q,p_j\|)\leq f(\|q,p_k\|,\|q,p_l\|)$. Thus the lemma is proved.
\end{proof}  \qed

\begin{lem} By using the higher order Voronoi partition and the distance function defined above, the performance function can be further transformed as
\begin{equation}
\label{double min achieve}
\begin{split}
\mathcal{H}=&\int_Q \min_{(i,j)\in C}f(\|q,p_i\|,\|q,p_j\|) \phi(q) dq\\
=&\sum_{\forall\mathcal{T}_{ij}\subset\mathcal{S}}\int_{V_{\mathcal{T}_{ij}}}f(\|q,p_i\|,\|q,p_j\|) \phi(q) dq\\
\end{split}
\end{equation}
\end{lem}
This lemma is a straightforward consequence of Lemma \ref{lemma1}.

According to Lemma \ref{lemma1}, as long as the three properties of $f(\cdot,\cdot)$ hold, one can transform the original performance function \eqref{double cover} into \eqref{double min achieve} by using the order-2 Voronoi concept.
Different $f(\cdot,\cdot)$ in the performance function will affect directly the optimization of the performance function, and thus the final configuration of sensors' positions.
The choice of $f(\cdot,\cdot)$ depends on specific requirements for different applications. Here are some typical $f(\cdot,\cdot)$, expressed in terms of norm.
\begin{enumerate}
  \item When $f(\cdot,\cdot)=\|q,p_i\|+\|q,p_j\|$, the sensing performance for each $q\in V_{\mathcal{T}_{ij}}$ is related to the distance sum from $q$ to two sites $p_i$ and $p_j$.  A more detailed explanation of an application will be discussed later in Section \ref{Real world applications}.
  \item When $f(\cdot,\cdot)=\|q,p_i\|^2+\|q,p_j\|^2$, the sensing performance is expressed by the squared distance sum to two sites $p_i$ and $p_j$.  In this case, as we are going to show in the next section, the controller expression has a strong relationship with the centroid of each cell.
  \item Suppose $f(\cdot,\cdot)=(\|q,p_i\|^n+\|q,p_j\|^n)^{1/n}$. As $n\rightarrow \infty$, there holds $f(\cdot,\cdot)=\max\{\|q,p_i\|,\|q,p_j\|\}$.
\end{enumerate}
The three examples just listed correspond to the order 2 case but the idea can be generalized to higher order cases. In the case of an order $k$ problem, the performance function $f$ is defined to have $k$ arguments. In addition, $f$ must be increasing in all its arguments, and symmetric in all its arguments. For example, an $L^p$ norm ($p>1$) of the $k$ arguments is a valid performance function. 
For the generalized performance function \eqref{double cover}, the optimal positions of sensors are $f$-dependent. In Section \ref{Sec:application}, we will provide more discussions on specific applications using different  $f(\cdot,\cdot)$.

\subsection{Controller design}

\subsubsection{Gradient-based controller}
\label{Gradient-based controller}
In order to minimize the performance function \eqref{double min achieve}, we can design a controller for each sensor with position $p_i$ as
$$ \dot p_i=-\frac{\partial \mathcal{H}}{\partial {p_i}}$$
The mobile sensor system with the above controller defines a gradient flow of the performance function \eqref{double cover}. According to the property of  gradient systems \cite{absil2006stable}, the above gradient controller provides a natural choice to optimize the performance function.
Furthermore, in order to implement the above controller, each agent needs to know the position of its neighbouring agents.
The neighbouring agents of $i$ are $\mathcal{N}_i = \{j|\mathcal{T}_{ij}\in \mathcal{P}\}\cup\{k,l|V_{\mathcal{T}_{ij}}\cap V_{\mathcal{T}_{jk}} \neq \emptyset \}$, which are agents that monitor the cells $V_{\mathcal{T}_{ij}}$ and the cells with common boundaries with these $V_{\mathcal{T}_{ij}}$.

\subsubsection{Cancellation of boundary terms}
In the following, we will present an explicit formula for the controller. In our order 2 problem, for each $p_i\in\mathcal{S}$, we are going to show
\begin{equation}\label{pupp}
%\begin{split}
\frac{\partial \mathcal{H}}{\partial {p_i}}=\sum_{\forall \mathcal{T}_{ij},p_i\in\mathcal{T}_{ij}}
\int_{V_{\mathcal{T}_{ij}}}\frac{\partial}{\partial {p_i}}f(\|q,p_i\|,\|q,p_j\|) \phi(q) dq
%\end{split}
\end{equation}

In the expression of $\mathcal{H}$ as shown in \eqref{double min achieve}, the domain of integration is a function of $p_i$. As a result, when one calculates the partial derivative of $\mathcal{H}$ with respect to $p_i$, one needs to deal with the problem of differentiation under the integral sign. Some basic facts about the problem of differentiating under the integral sign (where integrand and integration limits are functions of a parameter) can be found in \cite{flanders1973differentiation}.

\begin{figure}[!ht]
\begin{center}
\includegraphics[width=0.9\columnwidth]{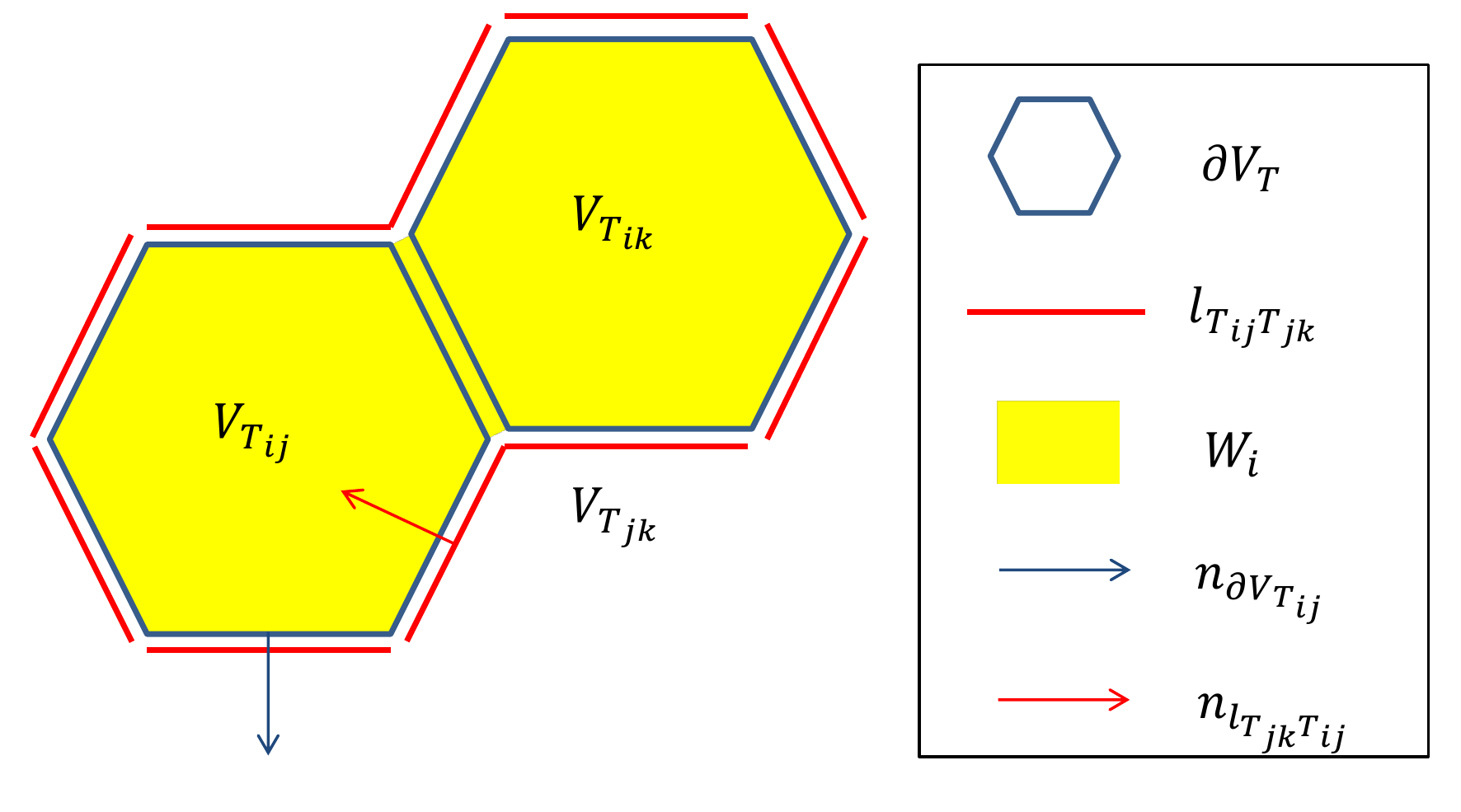}
\end{center}
\caption{Graphical representations of notations referred to each cell and the boundary}
\label{fig_notation}
\end{figure}

 One might expect that the right side of the above equation \eqref{pupp} should contain an additional term reflecting the dependence of the region $V_{\mathcal T_{ij}}$ on $p_i$; in effect, \eqref{pupp} is equivalent to the fact that this additional term evaluates as zero. The same phenomenon is well understood in the literature on first order Voronoi optimization problems, and an explanation in terms of elementary calculus concepts can be found in \cite{asami1991note}, which points out that it is a particular case of what is sometimes called the \textit{envelope theorem}: if a function $h(x,y)$ of two scalar variables is $C^2$ and $H(x)=\min_yh(x,y)$, then a simple calculation shows there holds
\begin{equation}\label{eq:envelope}
\frac{dH}{dx}=\frac{\partial h}{\partial x}
\end{equation}
when $y$ is set to the minimizing value associated with the particular $x$. This equation states that the tangent to an envelope curve (defined by $H(\cdot)$) at a particular point on the curve is also tangent to the particular parametrized curve $h(\cdot,y)$, $y$ being the parameter, passing through the same point $x$, and it extends trivially to the case of vector $x,y$.  It is also known that the $C^2$ dependence of $h$ on $y$ is not actually necessary, but this is more technical. For our application, one can think of $x$ as like the sensor positions (generators) and $y$ as like the boundary points of the cells. Note that the ordinary derivative in \eqref{eq:envelope} is replaced by the partial derivative on the left of \eqref{pupp} since $x$ is replaced by the set of $p_i$.

Analogously, there is also a relation between the Hessians of $H$ and $h$ when $y$ is set at the minimizing value associated with a particular $x$; call this value $y^*(x)$. A calculation, not very difficult, reveals that in the vector case
\begin{equation}
\nabla^2 H=\left[\begin{array}{c}
I\\\frac{\partial y^*(x)}{\partial x}\end{array}\right]^{\top}\left[\begin{array}{cc}
\frac{\partial ^2h(x,y)}{\partial x^2}&\frac{\partial ^2h(x,y)}{\partial x\partial y}\\
\frac{\partial ^2h(x,y)}{\partial x\partial y}&\frac{\partial ^2(h(x,y)}{\partial y^2}\end{array}\right]\left[\begin{array}{c}
I\\\frac{\partial y^*(x)}{\partial x}\end{array}\right]
\end{equation}
So nonnegativity or positive definiteness of the Hessian of $h$ maps into the same property for $H$. If we separately optimize the sensor positions and the cell boundaries and achieve a minimum, then we will also have a minimum if we assume cell boundaries are those appropriate to particular sensor positions, and just minimize with respect to the latter.

Now we are going to define some quantities in relation to order 2 Voronoi cells as shown in Figure \ref{fig_notation}. Let $\partial V_{\mathcal{T}_{ij}}$ and $\partial Q$ be the boundary of $V_{\mathcal{T}_{ij}}$ and $Q$ respectively. Further suppose that $l_{\mathcal{T}_{ij}\mathcal{T}_{jk}}=V_{\mathcal{T}_{ij}}\cap V_{\mathcal{T}_{jk}}$ is the common boundary of $V_{\mathcal{T}_{ij}}$ and $V_{\mathcal{T}_{jk}}$. Let $q_{\partial V_{\mathcal{T}_{ij}}}$ be a point on the boundary of $V_{\mathcal{T}_{ij}}$, and $q_{l_{\mathcal{T}_{ij}\mathcal{T}_{jk}}}$ be a point on the common boundary of $V_{\mathcal{T}_{ij}}$ and $V_{\mathcal{T}_{jk}}$. We also define $n_{\partial V_{\mathcal{T}_{ij}}}$ as the outward facing unit normal vector of $\partial V_{\mathcal{T}_{ij}}$, and $n_{l_{\mathcal{T}_{ij}\mathcal{T}_{jk}}}$ as the unit normal vector of $l_{\mathcal{T}_{ij}\mathcal{T}_{jk}}$ from cell $\mathcal{T}_{ij}$ to $\mathcal{T}_{jk}$.

In our problem, if we calculate the partial derivatives of $\mathcal{H}$ with respect to each sensor position $p_i$ under the integral sign, we have

\begin{equation}\label{puppder}
\begin{split}
\frac{\partial \mathcal{H}}{\partial {p_i}}&=\sum_{\mathcal{T}_{ij}\in \mathcal{P}_i}
\int_{V_{\mathcal{T}_{ij}}}\frac{\partial}{\partial {p_i}}f(\cdot,\cdot) \phi(q) dq\\
&+\sum_{\mathcal{T}_{ij}\in \mathcal{P}_i}
\int_{\partial Q \cap \partial V_{\mathcal{T}_{ij}}} f(\cdot,\cdot) \phi(q) \frac{\partial q_{ \partial V_{\mathcal{T}_{ij}}}}{\partial p_i} n_{\partial V_{\mathcal{T}_{ij}}} dq\\
&+\sum_{\mathcal{T}_{ij}\in \mathcal{P}_i}
\int_{\partial V_{\mathcal{T}_{ij}} \backslash \partial Q} f(\cdot,\cdot) \phi(q) \frac{\partial q_{\partial V_{\mathcal{T}_{ij}}}}{\partial p_i} n_{\partial V_{\mathcal{T}_{ij}}} dq\\
&+\sum_{\underset{V_{\mathcal{T}_{ij}}\cap V_{\mathcal{T}_{jk}}\neq \emptyset}{\mathcal{T}_{ij}\in \mathcal{P}_i,\mathcal{T}_{jk}\not\in \mathcal{P}_i}}
\int_{l_{\mathcal{T}_{ij}\mathcal{T}_{jk}}} f(\cdot,\cdot) \phi(q) \frac{\partial q_{l_{\mathcal{T}_{ij}\mathcal{T}_{jk}}}}{\partial p_i} n_{l_{\mathcal{T}_{jk}\mathcal{T}_{ij}}} dq\\
\end{split}
\end{equation}

Note the second line in \eqref{puppder} is always zero because $\partial Q$ is always stationary and its partial derivatives with respect to the entries of $p_i$ are always zero. By the definition of higher order Voronoi cell \cite{shamos1975closest}, we know $l_{\mathcal{T}_{ij}\mathcal{T}_{jk}}=l_{\mathcal{T}_{jk}\mathcal{T}_{ij}}$ and $n_{l_{\mathcal{T}_{ij}\mathcal{T}_{jk}}}=-n_{l_{\mathcal{T}_{jk}\mathcal{T}_{ij}}}$. As a result, the third and fourth lines in \eqref{puppder} cancel out with each other. Therefore \eqref{pupp} holds. Note that the results hold for general function $f(\cdot,\cdot)$ with the given properties.

\subsection{Higher order centroidal Voronoi partitions}

For any given $f(\|q,p_i\|,\|q,p_j\|)$, it is not easy to find the general relationship of \eqref{pupp} with cell centroids of $W_i$. However, motivated by the first order case, we  consider the case when
$f(\|q,p_i\|,\|q,p_j\|)=\frac{1}{2}(\|q,p_i\|^2 + \|q,p_j\|^2)$. In this case, the performance function becomes \begin{equation}\label{eq:highercentroidal}
\mathcal{H}(p_1, p_2,\cdots,p_n)=\int_Q \min_{(i,j)\in C} \frac{1}{2}(\|q,p_i\|^2 + \|q,p_j\|^2) \phi(q) dq
\end{equation}

Note the centroid and mass of a Voronoi cell $V_{\mathcal{T}_{ij}}$ are
$C_{V_{\mathcal{T}_{ij}}}=\frac{\int_{V_{\mathcal{T}_{ij}}}q\phi(q)dq}{\int_{V_{\mathcal{T}_{ij}}}\phi(q)dq}$
and ${M}_{V_{\mathcal{T}_{ij}}}=\int_{V_{\mathcal{T}_{ij}}}\phi(q)dq$ respectively.
In this case,
\begin{equation}\label{puppcentroid}
\begin{split}
\frac{\partial \mathcal{H}}{\partial {p_i}}&=\sum_{\mathcal{T}_{ij}\in \mathcal{P}_i}
\int_{V_{\mathcal{T}_{ij}}}\frac{\partial}{\partial {p_i}}\frac{1}{2}(\|q,p_i\|^2 + \|q,p_j\|^2) \phi(q) dq\\
&=-\sum_{\mathcal{T}_{ij}\in \mathcal{P}_i}\int_{V_{\mathcal{T}_{ij}}}(q-p_i)\phi(q)dq\\
&=\sum_{\mathcal{T}_{ij}\in \mathcal{P}_i}-{M}_{V_{\mathcal{T}_{ij}}}(C_{V_{\mathcal{T}_{ij}}}-p_i)\\
\end{split}
\end{equation}

Suppose further that the centroid and mass of $W_i$ are
$C_{W_i}=\frac{\sum_{\mathcal{T}_{ij}\in \mathcal{P}_i}{M}_{V_{\mathcal{T}_{ij}}}C_{V_{\mathcal{T}_{ij}}}}{\sum_{\mathcal{T}_{ij}\in \mathcal{P}_i}{M}_{V_{\mathcal{T}_{ij}}}}$
and ${M}_{W_i}=\sum_{\mathcal{T}_{ij}\in \mathcal{P}_i}{M}_{V_{\mathcal{T}_{ij}}}$ respectively.
Now we have
\begin{equation}\label{puppcentroidW}
\begin{split}
\frac{\partial \mathcal{H}}{\partial {p_i}}&=\sum_{\mathcal{T}_{ij}\in \mathcal{P}_i}-{M}_{V_{\mathcal{T}_{ij}}}(C_{V_{\mathcal{T}_{ij}}}-p_i)\\
&=-C_{W_i}\sum_{\mathcal{T}_{ij}\in \mathcal{P}_i}{M}_{V_{\mathcal{T}_{ij}}}+p_i\sum_{\mathcal{T}_{ij}\in \mathcal{P}_i}{M}_{V_{\mathcal{T}_{ij}}}\\
&=-{M}_{W_i}(C_{W_i}-p_i)\\
\end{split}
\end{equation}
The above result indicates that if $p_i$ moves towards the centroid of $W_i$ then $\mathcal{H}$ will decrease over time.
Note that this centroid in general moves when the sensors move, since the Voronoi diagram will change with moving sensors.
This result is very similar to the order 1 centroid coverage control case when $f(x)$ is defined as $f(x) = \frac{1}{2}\|x\|^2$.
The notation of $W_i$ corresponds to the order 2 case but   it can be generalized to higher order cases. In the case of an order $k$ problem, $W_i$   consists of all the regions currently monitored by agent $i$.
%In order 2 or higher order case, $W_i$ replaced the position of $V_i$ in the order 1 case.
Now we summarize the convergence results concerning the above gradient system as follows.

\begin{lem}\label{higherordercentroidal}
For a group of mobile agents with the closed-loop system induced by  \eqref{pupp}, all the agents' positions  will  converge to
the set of critical points of $\mathcal{H}$. Furthermore, by taking $f(\|q,p_i\|,\|q,p_j\|)=\frac{1}{2}(\|q,p_i\|^2 + \|q,p_j\|^2)$ and designing the controller as $\dot p_i =\alpha(C_{W_i}-p_i)$ where $\alpha$ is a positive gain, agents' locations will converge to the cell centroids which result in a higher order  centroidal  Voronoi configuration.
\end{lem}

The proof follows directly from the properties of gradient systems (see e.g. \cite{absil2006stable}) and is omitted here.

\begin{rem}
For general forms of $f$, analytical results is in general unavailable. However, one can still carry out the distributed control algorithm by computing a first-order  difference approximation to the gradient of $\mathcal{H}$.
\end{rem}

\subsection {Lloyd algorithm}

As for the first order case, a Lloyd algorithm is available for the case where $f(\|q,p_i\|,\|q,p_j\|)=\frac{1}{2}(\|q,p_i\|^2 + \|q,p_j\|^2)$, which as we just established leads to centroidal higher order Voronoi partitions.

With an arbitrary initial set of positions for the sensors, determine the associated higher order Voronoi partition, including the sets $W_i$ and then the centroids of these sets. Adopt the centroids as the new generators (positions for the sensors). Determine with these new generators a new higher order Voronoi partition and new sets $W_i$, and then the centroids of these sets, and so on.  One cycle of the algorithm corresponds to one update of the generator positions.

This is the higher order Lloyd algorithm. Then we can state;

\begin{prop}
Consider the performance function defined in \eqref{eq:highercentroidal}, and suppose a higher order Lloyd algorithm is used for the purpose of finding the optimum $p_i$ minimizing the index. Then execution of one step of the algorithm ensures that $\mathcal H(p_1,p_2,\dots,p_n)$ decreases unless the $p_i$ coincide with the centroids of their associated Voronoi regions.
\end{prop}

\textbf{Proof.} Observe that (using the notation introduced in Subsections \ref{order k voronoi partition} and \ref{Performance function and its relationship with generalized Voronoi partition})
\begin{eqnarray}
&&\mathcal H(p_1,p_2,\dots,p_n)\\\nonumber
&=&\sum_{\mathcal T_{ij}\subset \mathcal S}\int_{V_{\mathcal T_{ij}}}\frac{1}{2}(||q,p_i||^2+||q,p_j||^2)\phi(q)dq\\\nonumber
&=&\sum_{\mathcal T_{ij}\subset \mathcal S}[\frac{1}{2}M_{V_{\mathcal T_{ij}}}||p_i-C_{V_{\mathcal T_{ij}}}||^2+\frac{1}{2}M_{V_{\mathcal T_{ij}}}||p_j-C_{V_{\mathcal T_{ij}}}||^2\\\nonumber&&+J_{V_{\mathcal T_{ij}}}]\\\nonumber
\end{eqnarray}
where $J_{V_{\mathcal T_{ij}}}$ denotes the polar moment of inertia around the centroid of the region $V_{\mathcal T_{ij}}$. Now $\mathcal H$ reduces its value if $p_i$ is replaced by $p_i'$ such that
\begin{equation}
\sum_{\mathcal T_{ij}\subset \mathcal P_i}\frac{1}{2}M_{V_{\mathcal T_{ij}}}||p_i'-C_{V_{\mathcal T_{ij}}}||^2\leq \sum_{\mathcal T_{ij}\subset \mathcal P_i}\frac{1}{2}M_{V_{\mathcal T_{ij}}}||p_i-C_{V_{\mathcal T_{ij}}}||^2
\end{equation}
Observe that
\begin{eqnarray}
&&\sum_{\mathcal T_{ij}\subset \mathcal P_i}M_{V_{\mathcal T_{ij}}}||p_i'-C_{V_{\mathcal T_{ij}}}||^2\\\nonumber
&=&\sum_{\mathcal T_{ij}\subset \mathcal P_i}M_{V_{\mathcal T_{ij}}}||p_i'||^2-2\sum_{\mathcal T_{ij}\subset \mathcal P_i} M_{V_{\mathcal T_{ij}}} C_{V_{\mathcal T_{ij}}}^{\top}p_i'\\\nonumber
&&+\sum_{\mathcal T_{ij}\subset \mathcal P_i}M_{V_{\mathcal T_{ij}}}||
C_{V_{\mathcal T_{ij}}}||^2\\\nonumber
&=&M_{W_i}||p_i'||^2-2M_{W_i}C_{W_i}^{\top}p_i'+D
\end{eqnarray}
where $\partial{D}/\partial{p'_i}=0$. It is now trivial to observe that the choice
\begin{equation}
p_i'=C_{V_i}
\end{equation}
achieves the maximum reduction in $\mathcal H$. Moreover, reduction is always achieved unless the $p_i$ are already located at the cell centroids.  The update from $p_i$ to $p_i'=C_{\mathcal V_i}$  is precisely the choice delivered by one cycle of the Lloyd algorithm.  \qed

The above Lloyd algorithm only employs the gradient information, while for a faster convergence one may consider use of higher-order optimization methods (e.g. Newton-like algorithms). The application of Newton-like algorithm requires the performance function $\mathcal H$ defined in \eqref{double cover} to be a $C^2$ function. We conjecture that, by assuming a $C^2$ density function $\phi$, the quadratic distance function $f$, and the convex region $Q$, the performance function  $\mathcal H$ is $C^2$, by following the proof in   \cite{liu2009} which applies in the first order case. This conjecture is supported by the fact that the higher-order partition $V_{\mathcal{T}}$ shares the same properties as the order 1 partition $V_i$, as discussed in Subsection \ref{order k voronoi partition}. Such a conjecture, if true, also guarantees that the Lloyd map is singly differentiable, a fact which assists examination of the stability of fixed points of the map; stability properties of a fixed point can be characterized in terms of the eigenvalues of the derivative of the map at the fixed point.
Since this $C^2$ property is not the main focus of this paper, we will not dwell on this point but will consider it in more detail in  future research.

\subsection{The minimum sensing radius $k$-th order problem}

For a given number of sensors and a requirement that every point in a region be covered by at least $k$ sensors, this problem seeks to find an algorithm for positioning the sensors so that the common sensing radius of all the sensors can be minimized. Here, `coverage' of a point by a sensor corresponds to the point lying closer to the sensor than the sensing radius. Minimization of the sensing radius will often be a desirable objective because it will minimize power usage by the sensors.  One applications context for this problem is the task of localizing targets using range-only sensing. Generically, a target must be seen by three sensors, since with two sensors alone, the target position is normally subject to a binary ambiguity. Various aspects of this problem are treated in e.g.\cite{6258029,So2005}.

In addition, the reference \cite{So2005} concludes that a stationary point of the relevant performance function is defined by a choice of sensor positions $p_i, i=1,2,\dots,n$ such that $p_i$ is the Chebyshev center for $\mathcal W_i$, which was defined as the union of the $k$-th order Voronoi cells for which  $p_i$ is one of the generating set of $k$ generators. As in the order 1 case, an iterative algorithm based on taking a set of sensor positions, finding the associated $k$-th order $\mathcal W_i$, and then adopting new sensor positions as the Chebyshev centers of the $\mathcal W_i$.

We remark that, in comparison with the order 1 problem, the regions $W_i$ for a $k\geq 2$ order case are not necessarily convex, but the Chebyshev center (defined by the minimal radius circumcenter) is still unique and depends continuously but not everywhere differentially on the generator positions. The simulation result for the continuous Chebyshev order 2 case (where agent $i$ moves towards the Chebyshev center of region $W_i$, with velocity proportional to its displacement from the Chebyshev center) is very similar to that of quadratic cost \eqref{eq:highercentroidal}, except that an optimal sensor position could be at the boundary of the overall region $Q$.  For this reason no separate simulations are described in the subsequent section treating simulations.  Trajectories for the continuous time algorithm are not especially different to those for the iterative algorithm of \cite{6258029}.

\subsection{Collision avoidance}\label{subsection:CA}
Due to the gradient-based nature of our controller, sometimes the obtained solution may be an undesired local optimum where some pair of agents are collocated. In order to avoid these undesired local optima, one can use modified controller functions to prevent collocation. One possible way is to modify the controller as $\dot{p}_i = -\frac{\partial \mathcal{H}}{\partial p_i} + u_i(\|p_i,p_j\|)$, where $u_i(\|p_i,p_j\|)$ is a control term to prevent the collocation (see e.g. \cite{hussein2007effective}). Another possible way is to modify the performance index $f(\cdot,\cdot)$. One might think of modifying it as $f=\frac{\|q,p_i\|+\|q,p_j\|}{\|p_i,p_j\|}$ for example. However, the  modification is not feasible because the $f$ function in this case does not satisfy the three properties listed earlier and thus the Voronoi diagram cannot be used to generate optimal partition.

We can observe that the two quantities $\|q,p_i\|$ and $\|q,p_j\|$ are the same $\forall q\in V(\mathcal{T}_{ij})$ if and only if the two agents $p_i$ and $p_j$ are collocated. Thus a possible way to avoid collision is to modify the function $f$ as
$$f=\|q,p_i\|^2+\|q,p_j\|^2-a\cdot abs(\|q,p_i\|^2-\|q,p_j\|^2)$$
where $abs(\cdot)$ denotes absolute value and $a$ is a positive real number. Note the function $f$ satisfies the three properties as long as $a\in(0,1]$.

Furthermore, if we let $a=1$, then the performance index function $f$ becomes $f=2\cdot \min\{\|q,p_i\|^2,\|q,p_j\|^2\}$, and the optimization problem degenerates to an order 1 coverage control problem.

\subsection{Convergence issues}

By and large, convergence issues for the gradient descent algorithm and the Lloyd algorithm in the $k$-th order case are similar to those in the order 1 case. Of course, it would be desirable if such algorithms converged to a global minimum. This however is not always guaranteed.

% Analysis of critical points of an index is of course often aided by knowing properties of the Hessian (second derivative). Thus a preliminary question that should be considered is whether this exists. For order 1 problems, convexity of the region $Q$ assures this property, \cite{liu2009}. {\color{red} Zhiyong is still investigating the $C^2$ property for $k$-th order problems . Some of the following text refers} If the region is nonconvex, as shown in \cite{liu2009} there are nongeneric problems where the Hessian is discontinuous. Since in the $k$-th order problem, the individual cells (sets we have labelled $W_i$)  are not convex, it might be that for nongeneric problems, discontinuity occurs even if $Q$ is a convex polytope. Such discontinuities may create problems in using algorithms.

We note that for order 1 problems, there exist centroidal Voronoi tessellations for which the Hessian is indefinite or singular. A tedious calculation demonstrates this fact for a rectangular region $Q$, with two cells obtained by joining midpoints on the shorter sides of the rectangle. Suppose the shorter side is of length 1. When the longer side has length exceeding $\sqrt\frac{3}{2}$, the Hessian associated with the centroidal tessellation will have one negative eigenvalue, and when the longer side has length equal to $\sqrt\frac{3}{2}$, the Hessian will have one zero eigenvalue.
%{\color{red} I will be asking Zhiyong to see if he can provide a figure. Zhiyong to Brian: I later realize that putting a figure here may cause divergence of the main topic--as the figure will be only for order 1 case which is not actually the focus of this section. If the above statement is self-explained we may not need a figure here. }

This is most easily seen by working with the Lloyd map $T$, mapping generators $p_1,p_2,\dots$ of a collection of Voronoi sets  to the centroids $c_1,c_2,\dots$ of those sets. The Lloyd map is known to be $C^1$ for a convex region $Q$.   At a centroidal tessellation, there holds $\nabla {\mathcal H}=0, T(p)=p$ and there further holds \cite{du1999centroidal}
\begin{equation}
\nabla^2\mathcal H=2M[I-\nabla T]
\end{equation}
where $M$ is a positive definite diagonal matrix involving the masses $M_{V_i}$ of the Voronoi regions. The eigenvalues of the Jacobian $\nabla T$ must be less than 1 in magnitude for the centroidal tessellation to be a stable equilibrium of the Lloyd algorithm, and these eigenvalues are relatively easy to compute for any centroidal tessellation, including the 2 cell tessellation just mentioned, and lead to the conclusion recorded above.

Of course, since the cost function in a steepest descent algorithm can never increase and is bounded below, in the limit as $t\rightarrow \infty$, this cost function necessarily goes to a limit. Further, $\dot{\mathcal H}$ can be checked for an order one problem to be
\begin{equation}
\dot{\mathcal H}=-\alpha \sum_iM_{V_i}||[T(p)]_i-p_i||^2
\end{equation}
Here $[T(p)]_i$ denotes the centroid of the Voronoi region $V_i$ and this equation is a consequence of the formula \eqref{eq:orderonemotion}. Because $T(p)$ is $C^1$ in $p$ and $\dot p$ is continuous in time and bounded, $\dot{\mathcal H}$ itself is differentiable with respect to time and then
by Barbalat's lemma, see e.g. \cite{S99}, $\dot{\mathcal H}$ itself tends to zero. This says that there is convergence to a centroidal Voronoi tessellation. If the associated Hessian happens to have a negative eigenvalue, the equilibrium will be a saddle point, and thus effectively unstable. From an arbitrary initial condition, one would not expect such an equilibrium to be reached. But from a thin set, it will normally be reachable. This argument will extend to the higher order case provided $T(p)$ is a $C^1$ function of $p$, using \eqref{puppcentroidW} and the law presented in Lemma \ref{higherordercentroidal}. There may be pathological cases where it is not, but these have not revealed themselves in simulations.

Finally, we remark that for certain sets $Q$ which are not convex polygons, there may be a continuum of order 1 centroidal Voronoi tessellations. Suppose for example that $Q$ is a circle of radius 1, and that there are four generators. If these generators are equally spaced around a circle with the same center as $Q$ but with radius $\pi/4$, they will be generators of a centroidal Voronoi tessellation. Obviously, there is a continuum of possibilities.

%\begin{remark}
%In this section we mainly focus on generalized coverage control using an order 2 Voronoi partition. We mention that the coverage scheme can be extended to higher order Voronoi partitions by modifying the performance function stated in \eqref{double cover} with reasonable distance functions. Also, the main analysis for the 2-D case can be extended straightforwardly to 3-D space coverage. The controller for the 3-D case takes the same form as in \eqref{pupp} but the derivation requires a more sophisticated analysis of boundary issues.
%\end{remark}
%\begin{remark}
%The discretized version of the original order 1 coverage control problem in the field of data analysis and image processing is called k-mean clustering \cite{du1999centroidal}. To the best of our knowledge, there is no previous literature on the k-mean clustering problem corresponding to the `order 2' coverage problem in the continuous case. In the future, we intend to use the idea of higher order Voronoi partition in the research of k-mean clustering methods.
%\end{remark}

\section{Higher order coverage for discrete sets}   \label{Sec:higher_order}

Instead of a convex area $Q$ one can consider the coverage problem for a finite set of discrete points
$Q=\{q_1,\ldots, q_N\}$ in $\mathbb{R}^n$.
 This problem arises in applications where only a finite number of points of interest are
 to be observed by sensors which can move freely in the plane. First order coverage is related to the clustering problem, where a set of $N$ observations is partitioned
into into $m$ clusters according to some notion of similarity, see e.g. \cite{aggarwal2015}.
The Lloyd algorithm for first order coverage of discrete point sets is also known as the $k$-means
algorithm and widely applied for clustering problems \cite{Reddy2013}.
Higher order coverage can be viewed as a clustering approach where points are assigned to the multiple clusters simultaneously.
Assigning points to multiple clusters appears in fuzzy c-means clustering \cite{Dunn1973,Bezdek1981}.
However, in these approaches each data point is assigned to all clusters according to a fuzzy membership function.

The order 1 version of this problem is known as the clustering problem, and it attempts to partition $N$ observations into $m$ clusters, with each observation belonging to the cluster whose mean is nearest to the observation, see e.g. \cite{Ha75}.
The Lloyd algorithm presented earlier was advanced as a tool for solving this problem.
Here we are interested in the higher order clustering problem, where points are assigned the multiple clusters simultaneously.
Assigning points to multiple clusters appears in fuzzy c-means clustering \cite{Dunn1973,Bezdek1981},
which assigns each data point to all clusters according to a fuzzy membership function.

In the following we let $k$ denote the order that we consider; note $k\leq N$ is a finite number. 
It is straightforward to extend the generalized Voronoi partitions ${V}_\mathcal{T}$ to the discrete points case.
For technical reasons
we require that for different $\mathcal{T}$ and $\mathcal{T}'$ the sets ${V}_\mathcal{T}$, ${V}_\mathcal{T}'$ have empty intersection.
This can be ensured by assigning a $q_j$ only to the ${V}_\mathcal{T}$ where $\mathcal{T}$ is the smallest w.r.t. a suitable order of the $\mathcal{T}$,
e.g. a lexicographical order.
For a discrete data set the measure $\phi(q)dq$ in the performance function
is replaced by a discrete measure supported on the finite set $\{q_1,\ldots,q_N\}$.
This turns the integral into a finite sum which gives
the performance function
\begin{equation} \label{perf m-means}
\mathcal{H}(p_1,\ldots,p_m) = \sum_{l=1}^N w_l \min_{(i_1,\cdots,i_k)\in C}f(\|q_l,p_{i_1}\|,\|q_l,p_{i_2}\|,\cdots,\|q_l,p_{i_k}\|)
\end{equation}
with $w_1,\ldots, w_N$ suitable weights (which may be all the same).
Our goal is again to find minima of the performance function.
Any sensor placement yields an assignment of the $q_j$ to the  $W_i$, where analogously to the convex set case
a $W_i$ is the union of the cells (which are finite sets) assigned to sensor $p_i$.

A classical and well-known approach for achieving clustering in the first order case is the  so-called $k$-means algorithm  (The use of the symbol $k$ in the literature is common, and long-standing, see e.g. \cite{Ha75}. In terms of our notation, the term $m$-means algorithm is more appropriate.)

The Lloyd / $k$-means algorithm alternates between updating the estimates of centroids of the clusters and computing new clusters in each iteration until convergence is reached.
We extend this algorithm to our coverage problem, and we refer to this extension as the higher-order $m$-means
algorithm  (The use of the symbol $k$ in the literature is common, and long-standing, see e.g. \cite{aggarwal2015}. In terms of our notation, the term $m$-means algorithm is more appropriate.).
As an extension to the conditions on $f$ stated in Section \ref{order 2 section}, we assume that for all $p_{i_1},\cdots,p_{i_k}\in\mathbb{R}^n$
the function $q\mapsto f(\|q,p_{i_1}\|,\cdots \|q,p_{i_k}\|)$ is monotonically increasing, invariant w.r.t. the permutation of its arguments, and strictly convex.
We define for $p_1,\ldots,p_m \in\mathbb{R}^n$ and arbitrary non-empty subsets $W_1,\ldots,W_m\subset Q$
an extended performance function
\begin{multline*}
\hat{\mathcal{H}}(p_1,\ldots,p_m,W_1,\ldots,W_m) = \\
\sum_{\substack{i_1,\cdots,i_k=1\\ i_1\neq\cdots\neq i_k}}^m \sum_{q\in W_{i_1}\cap\cdots\cap W_{i_k}} w_q f(\|q,p_{i_1}\|,\|q,p_{i_2}\|,\cdots,\|q,p_{i_k}\|)
\end{multline*}
with $w_q$ the weight corresponding to $q\in Q$ in \eqref{perf m-means}.
In the first order covering case the Lloyd / $k$-means algorithm
can be viewed as alternating between optimization steps on the $p_i$ and the $W_i$ variables
\cite{selim1984}.
Thus for $p_1,\ldots,p_m\in \mathbb{R}^n$, $W_1,\ldots,W_m\subset Q$ we define $(C_1,\ldots,C_m)\in\mathbb{R}^{mn}$
by
\begin{multline*}
(C_1,\ldots,C_m) = \\
\operatorname*{argmin}_{\hat{p}_1,\ldots,\hat{p}_m\in\mathbb{R}^n} \sum_{\substack{i_1,\cdots,i_k=1\\ i_1\neq\cdots\neq i_k}}^m \sum_{q\in W_{i_1}\cap\cdots\cap W_{i_k}} w_q f(\|q,p_{i_1}\|\cdots,\|q,p_{i_k}\|)
\end{multline*}
Due to the convexity assumption on $f$ the $C_i$ are well-defined.
For the special case that $f(\|q,p_{i_1}\|,\cdots,\|q,p_{i_k}\|)=g(\|q,p_{i_1}\|)+\cdots+g(\|q,p_{i_k}\|)$ we have
\[
C_i =
\operatorname*{argmin}_{p\in\mathbb{R}^n} \sum_{q\in W_i} w_q g(\|q,p\|).
\]
In particular, for $g(\|q,p_i\|)=\|q,p_i\|^2$ and all $w_q=1$ the $C_i$ is the mean of the points in $W_i$.

{\bf Algorithm - second order $m$-means}
\begin{enumerate}
\item \label{step1}Start with  random values for the $p_i$.
Compute the corresponding $W_i$ where
\begin{align*}
W_i = \{ q_j\;|\; & j=1,\ldots,N ; \text{ there are at most } k-1  \\
&\text{ numbers of } p_l \text{ with}\\
&\|p_l-q_j\|<\|p_i-q_j\|\}  \}
\end{align*}
\item \label{step2}
For $i=1,\ldots,m$ do:
\begin{enumerate}
\item Compute the $C_i$.
\item Set $p_i= C_i$.
\end{enumerate}
\item \label{step3} Compute the new $W_i$ as in step \ref{step1}. If one $W_i$ is empty restart the algorithm with step \ref{step1}.
\item If $\hat{\mathcal{H}}$ is not decreased by the previous steps \ref{step2} or \ref{step3},
stop the algorithm. Otherwise go to step \ref{step2}.
\end{enumerate}

%We now discuss the convergence properties of the second order $m$-means algorithm.
%Consider the extended cost function
%\begin{multline*}
%\hat{\mathcal{H}}(p_1,\ldots,p_m,W_1,\ldots,W_m) = \\
%\sum_{\substack{i,j=1\\ i\neq j}}^m \sum_{q\in W_i\cap W_j} f(\|q,p_i\|,\|q,p_j\|)
%\end{multline*}

By construction %and Lemma \ref{lemma1}
each iteration in Step \ref{step2} does not increase $\hat{\mathcal{H}}$.
Further, Step \ref{step3} also does not increase $\hat{\mathcal{H}}$. Note that at this point
it is necessary that each $q_i$ is assigned only to two $p_j$. Due to boundedness of $\hat{\mathcal{H}}$ from below the values of $\hat{\mathcal{H}}$ converge.
If $f(\|q,p_i\|,\|q,p_j\|)=g(\|q,p_i\|)+g(\|q,p_j\|)$ each $C_i$
depends only on $W_i$. Further, there is only a finite number of possible values for each $W_i$
and the algorithm terminates when the cost is not decreasing. Thus for such performance
functions the algorithm stops after a finite number of steps.
\begin{prop} Denote by $p_i^l$, $W_j^l$ the iterates of the $k$th order $m$-means algorithm.
Then the sequence $\hat{\mathcal{H}}(p^l_1,\ldots,p^l_m,W^l_1,\ldots,W^l_m)$ converges.
For $f(\|q,p_{i_1}\|,\cdots,\|q,p_{i_k}\|)=g(\|q,p_{i_1}\|)+\cdots+g(\|q,p_{i_k}\|)$ the $m$-means algorithm terminates after a finite number of steps.
\end{prop}
Note that the argument for convergence after a finite number of steps is the same as for the standard $k$-means algorithm \cite{selim1984}.

\begin{figure}[!ht]
\subfigure[Evolution of the centroids]{
\includegraphics[width=0.44\columnwidth]{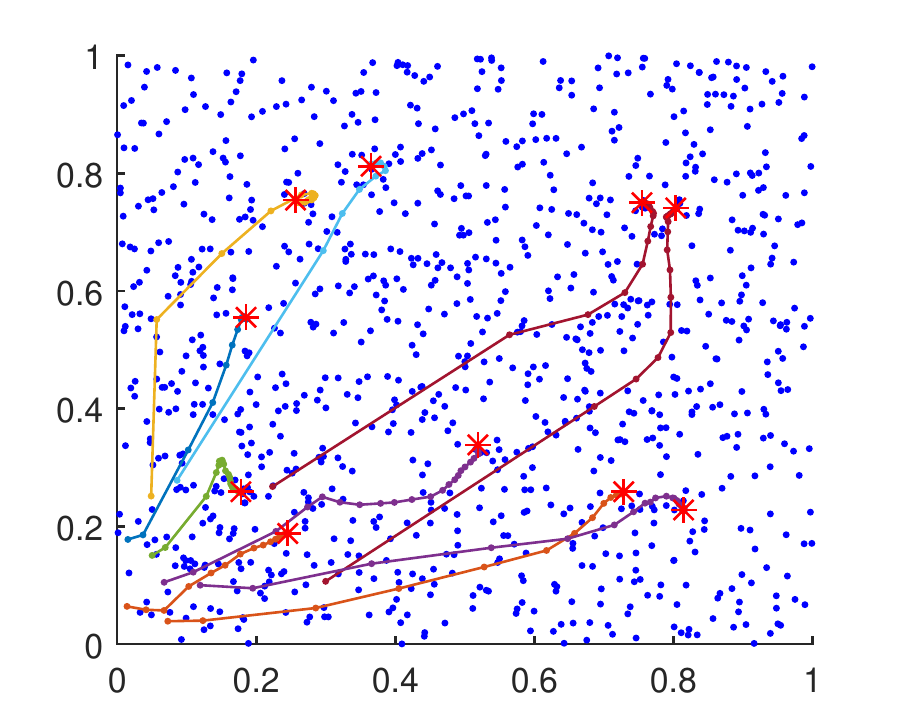}
\label{s11discrete}
}
\subfigure[Evolution of the performance function]{
\includegraphics[width=0.44\columnwidth]{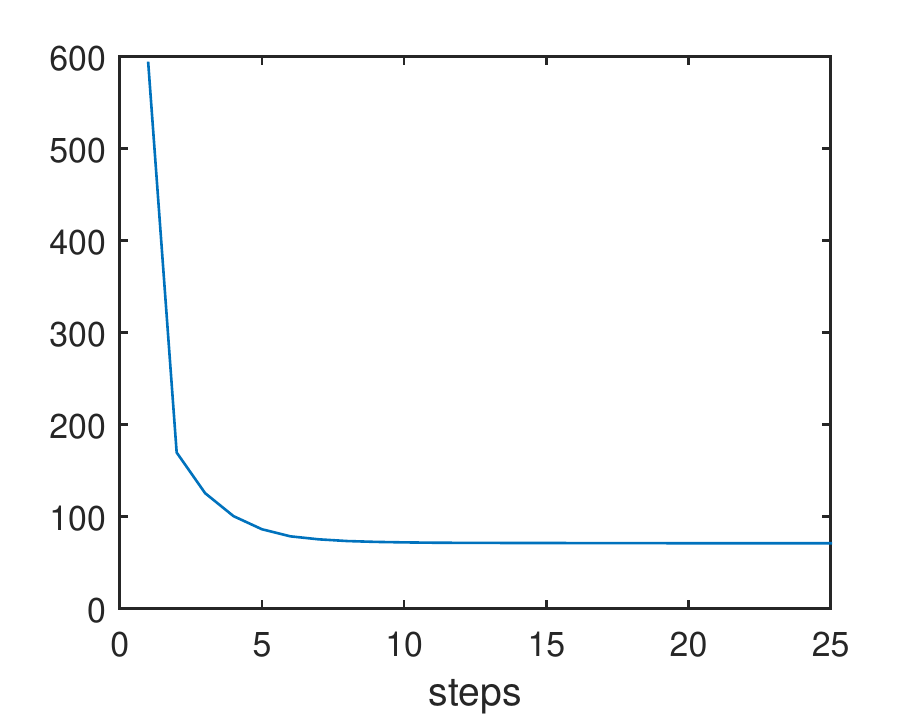}
\label{s12discrete}
}
\caption{Second order m-means}\label{second order k-means figure}
\end{figure}

%Note that this algorithm is effectively a form of Lloyd map, involving higher order partitioning and discrete data points.
Figure \ref{second order k-means figure} illustrates this behavior for 1000 uniformly distributed random data points in the square $[0,1]^2$. We use $m=10$ with random initial positions in the square $[0,0.3]^2$. Here we consider second order case. 
The performance function is monotonically decreasing and the algorithm terminates after a finite number of steps. However, it seems not to converge to a global minimum as the final positions (marked by red crosses) of some of the sensors are very close together.

Note that similarly to the $k$-means algorithm, the second (and higher order) $m$-means algorithms can be interpreted in the framework of the EM algorithm \cite{mclachlan2007algorithm}. Due to space limitations we will not pursue this direction in this paper.

\section{Simulation results and discussion}   \label{Sec:simulation}

\begin{figure}[!ht]
\subfigure[Initial sensor positions and order 2 Voronoi partition]{
\includegraphics[width=0.44\columnwidth]{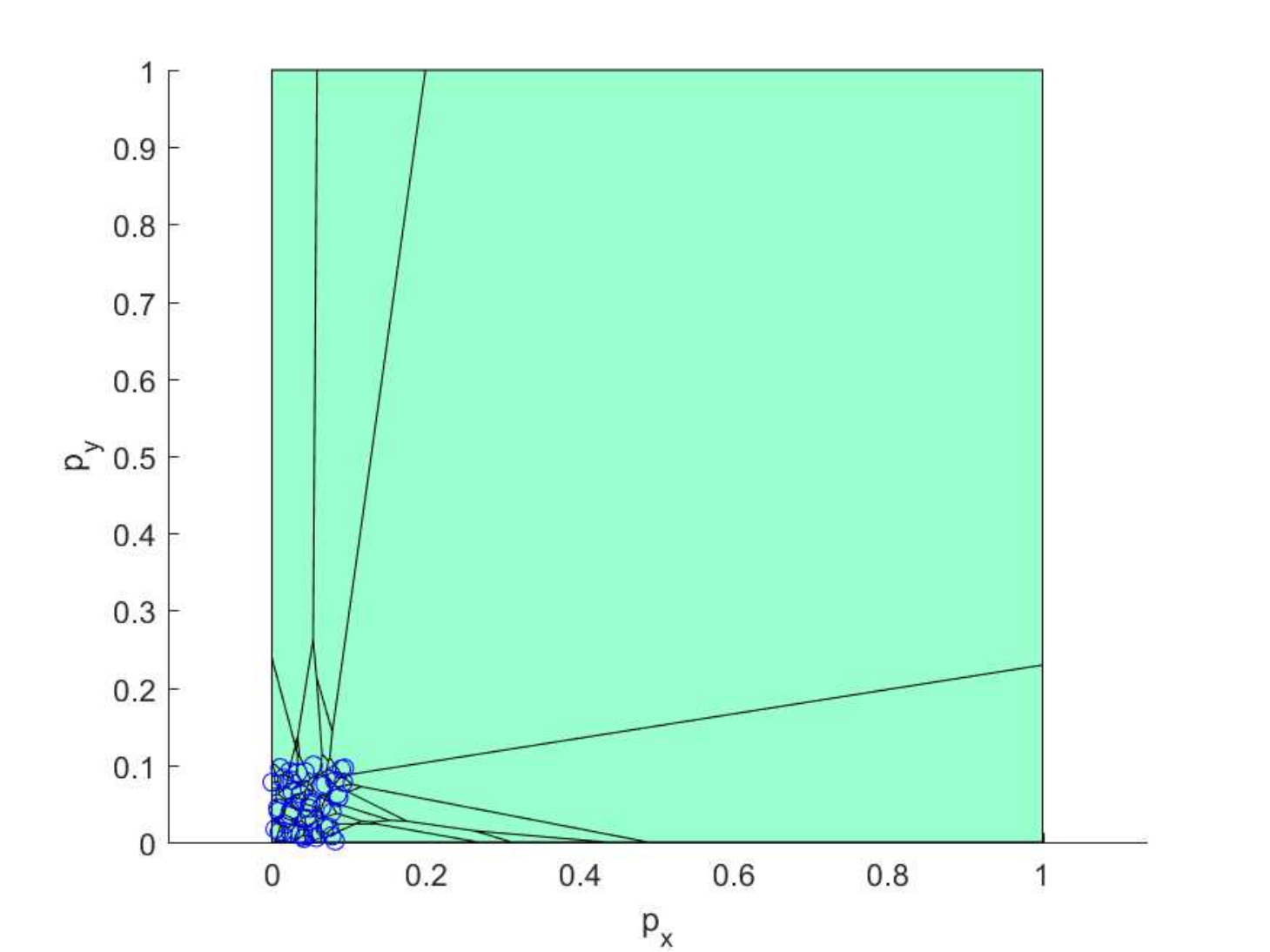}
\label{s11}
}
\subfigure[Moving trajectory of sensors]{
\includegraphics[width=0.44\columnwidth]{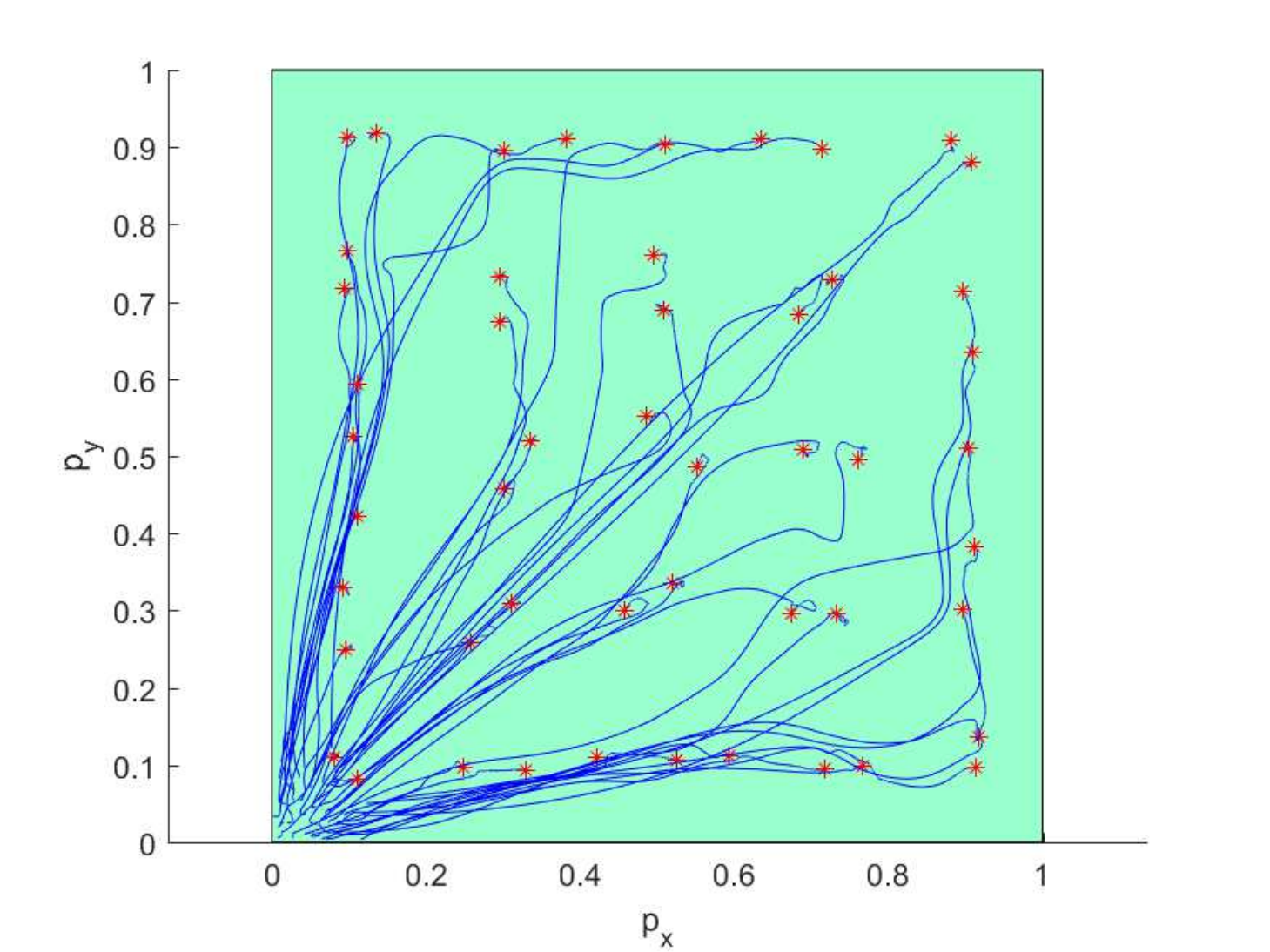}
\label{s12}
}
\subfigure[Final sensor positions and order 2 Voronoi partition]{
\includegraphics[width=0.44\columnwidth]{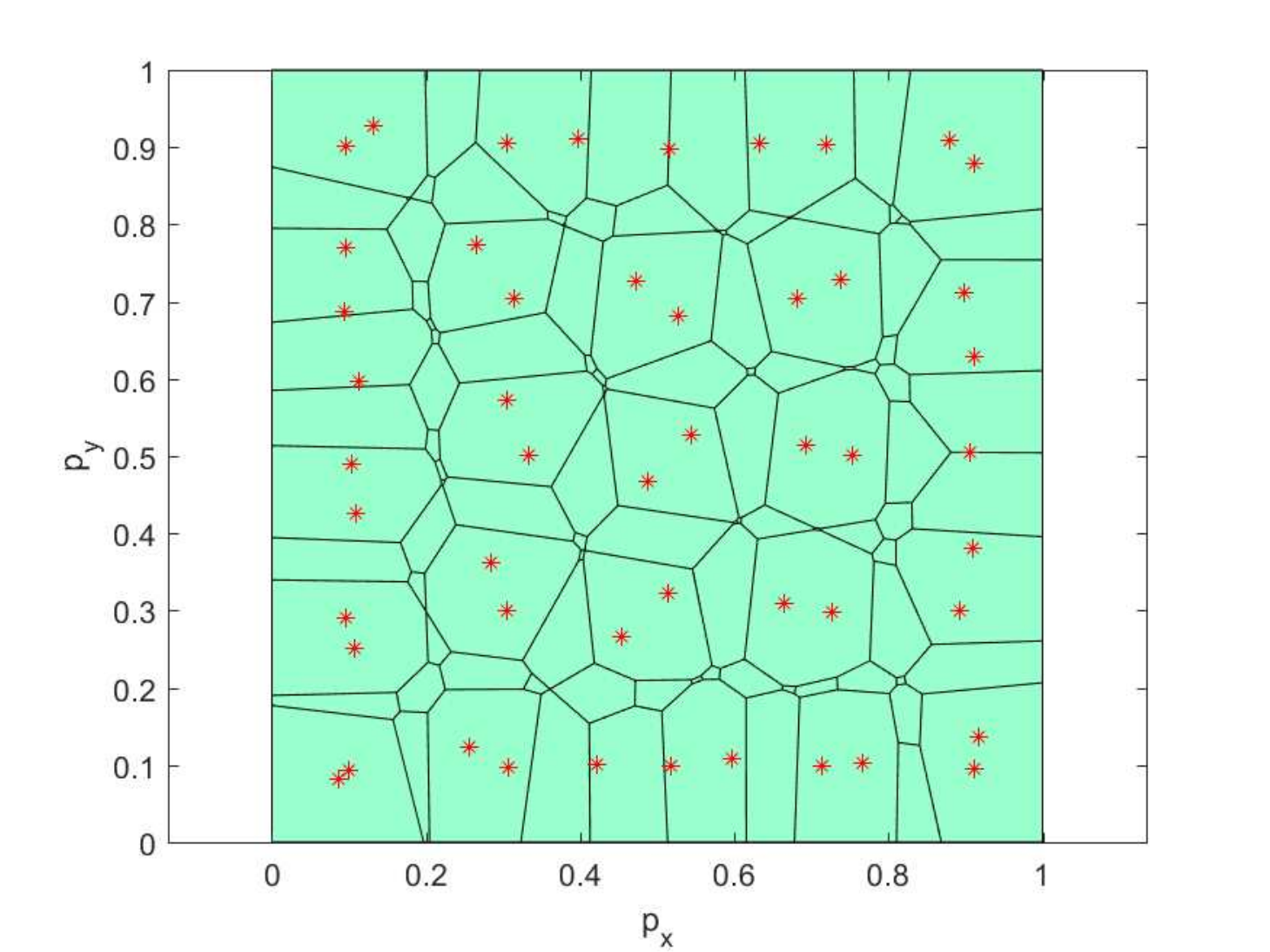}
\label{s13}
}
\subfigure[Evolution of the coverage objective function]{
\includegraphics[width=0.44\columnwidth]{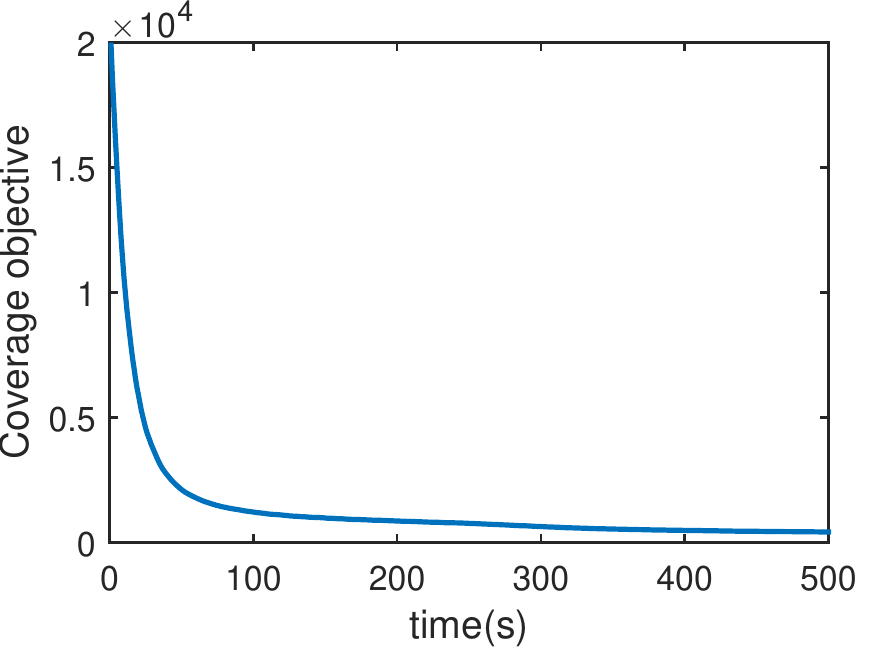}
\label{s14}
}
\caption{A simulation result with 50 mobile sensors} \label{50SensorResult}
\end{figure}

A simulation with 50 agents is shown in Figure \ref{50SensorResult}, in which we have chosen  $f(\cdot,\cdot)=\frac{1}{2}\|q,p_i\|^2+\|q,p_j\|^2$ and $\phi(q)=1$. 
Figure \ref{50SensorResult}(a) shows the initial positions of a group of agents.  Agents start randomly from the bottom-right corner. This figure also shows the order 2 Voronoi partition with these initial positions. Apart from that, Figure \ref{50SensorResult}(b) shows the moving trajectory of these agents using the controller designed in Section \ref{Gradient-based controller}.  Each  curve represents the trajectory of one agent and each star represents the final position of the agent. In addition, Figure \ref{50SensorResult}(c) shows the final positions of the group of agents and also the final order 2 Voronoi partitions. Furthermore, Figure \ref{50SensorResult}(d) shows the evolution of the value of the performance function from time $t=0$ to time $t=50$. As shown in this figure, the agent positions reach an equilibrium in the end.

\begin{figure}[!ht]
\subfigure[Moving trajectory of sensors]{
\includegraphics[width=0.46\columnwidth]{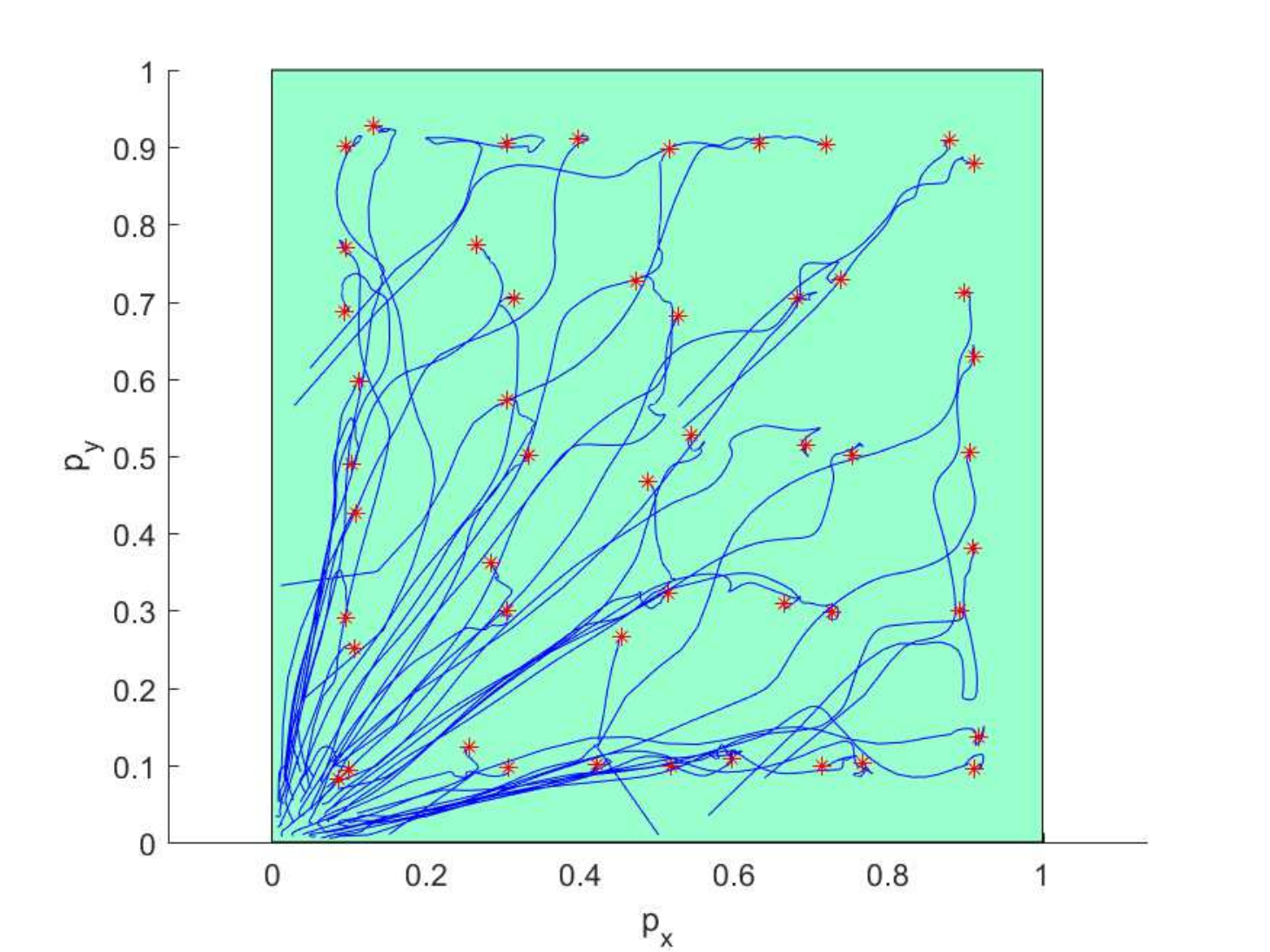}
\label{s21}
}
\subfigure[Final sensor positions and order 2 Voronoi partition]{
\includegraphics[width=0.46\columnwidth]{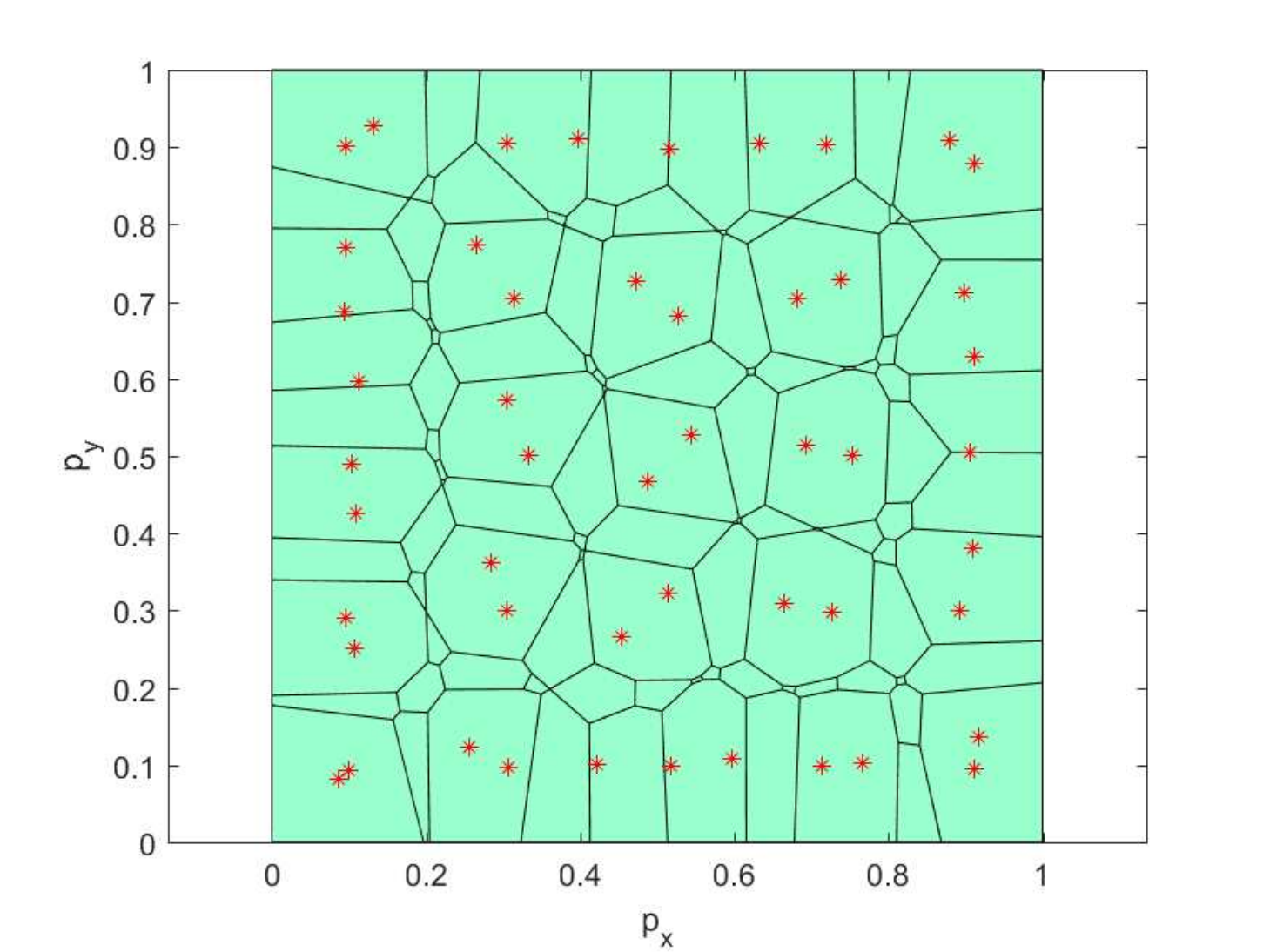}
\label{s22}
}
\caption{A simulation result of Lloyd algorithm} \label{50SensorResult_Lloyd}
\end{figure}

In addition, Figure \ref{50SensorResult_Lloyd} gives the result of an order two Lloyd algorithm with the same initial positions and cost function as for Figure \ref{50SensorResult}. Compared to the continuous counterparts, the final positions are very close (Figure \ref{50SensorResult} (c) vs. Figure \ref{50SensorResult_Lloyd} (b)) but the trajectories are not identical (Figure  \ref{50SensorResult} (b) vs. Figure \ref{50SensorResult_Lloyd} (a)).
Note that the controller designed in Section 3.2.1 is a continuous distributed control law, and so can be implemented on autonomous vehicles with limited maximal velocity. The Lloyd algorithm, on the other hand, is a fast-converging discrete version and is better for the purpose of computing the optimal positions.

%Due to the gradient-based nature of our controller, sometimes the obtained solution may be an undesired local optimum where some pair of agents are collocated. In order to avoid these undesired local optima, one can use modified $f(\cdot,\cdot)$ functions to prevent collocation. One possible modification is $$f(\cdot,\cdot)=\|q,p_i\|+\|q,p_j\|+g(\|p_i,p_j\|)$$ Note $g(\|p_i,p_j\|)$ goes to infinity when $\|p_i,p_j\|$ is very small and equals zero when $\|p_i,p_j\|$ is not so small.

Due to the gradient-based nature of our controller, sometimes the obtained solution may be an undesired local optimum where some pair of agents are collocated. In order to avoid these undesired local optima, one can use modified controller functions to prevent collocation. As noted previously, one can modify the controller as $\dot{p}_i = -\frac{\partial \mathcal{H}}{\partial p_i} + u_i(\|p_i,p_j\|)$, where $u_i(\|p_i,p_j\|)$ is a control term to prevent the collocation (see e.g. \cite{hussein2007effective} and Subsection \ref{subsection:CA}).
Such modifications also have much practical significance. For example, the performance of a bi-static radar (discussed further in the next section) is very poor when the transmitter is collocated with the receiver. Furthermore, for mobile agents, it is always desirable to keep a safe distance between each pair of agents so as to avoid collision.

\subsection{Regular shape in order 2 case}
The regular hexagon shape result for the order 1 case which often arises with a large number of agents cannot be easily generalized to order 2 case. However, as shown in the simulation in  \cite{JSA2015}, the order 2 centroidal optimal tessellation has some regular shape away from the boundaries of $Q$. In order to study possible regular shapes when the number of agents goes to infinity, we first introduce the notion of unit torus \cite{penrose2003random}. A unit torus in 2D is a square $[-\frac{1}{2},\frac{1}{2})$, in which the distance between two points $X$ and $Y$ are defined as
$$dist(X,Y)=\min \{\|X-Y+z\|:z\in \mathbb{Z}^2 \}$$

By using the unit torus, one can reduce the boundary effects to the optimal coverage shape. Simulation results are given on the unit torus as below.

\begin{figure}[!ht]
\subfigure[12 agents result]{
\includegraphics[width=0.46\columnwidth]{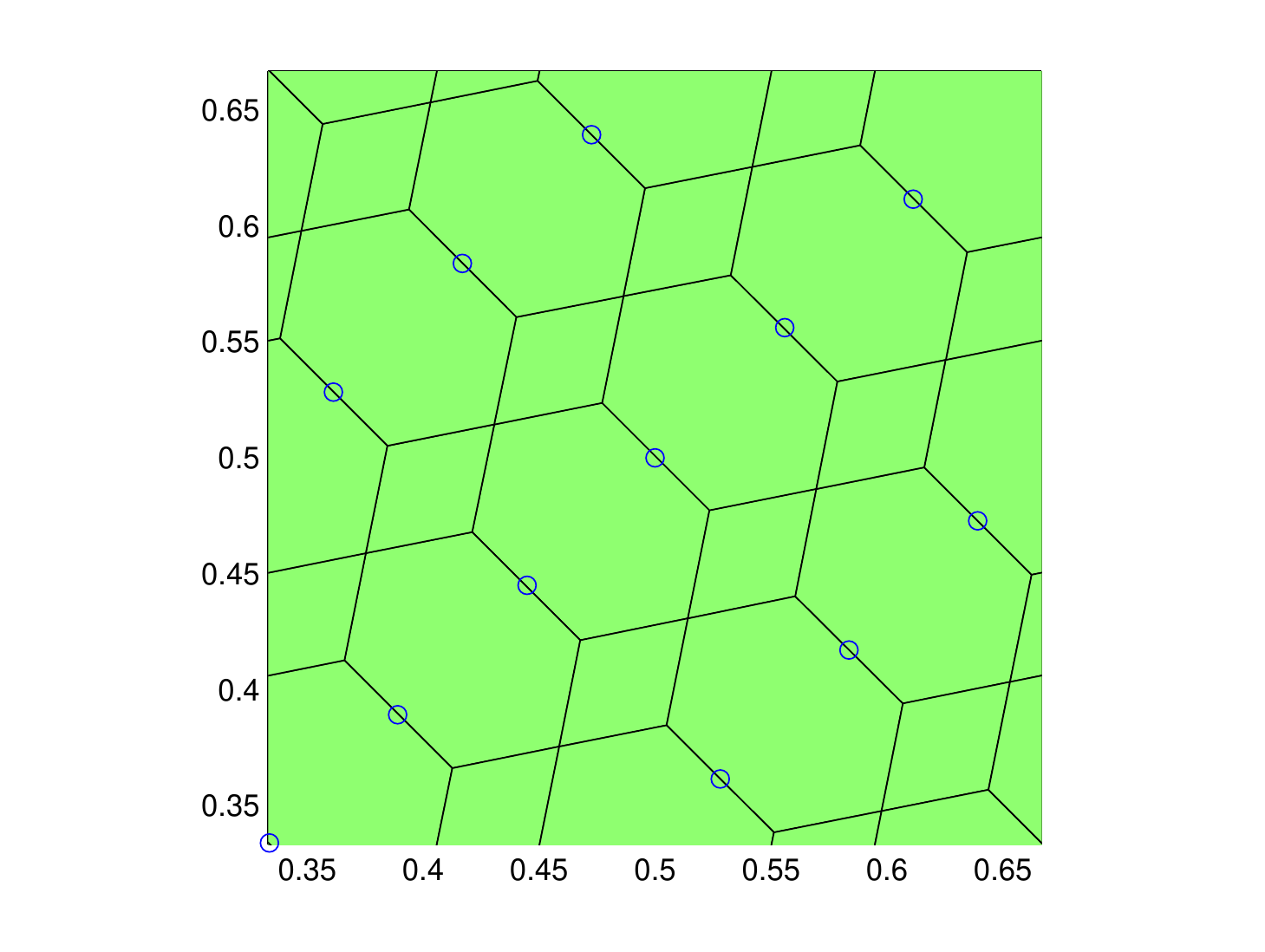}
}
\subfigure[36 agents result]{
\includegraphics[width=0.46\columnwidth]{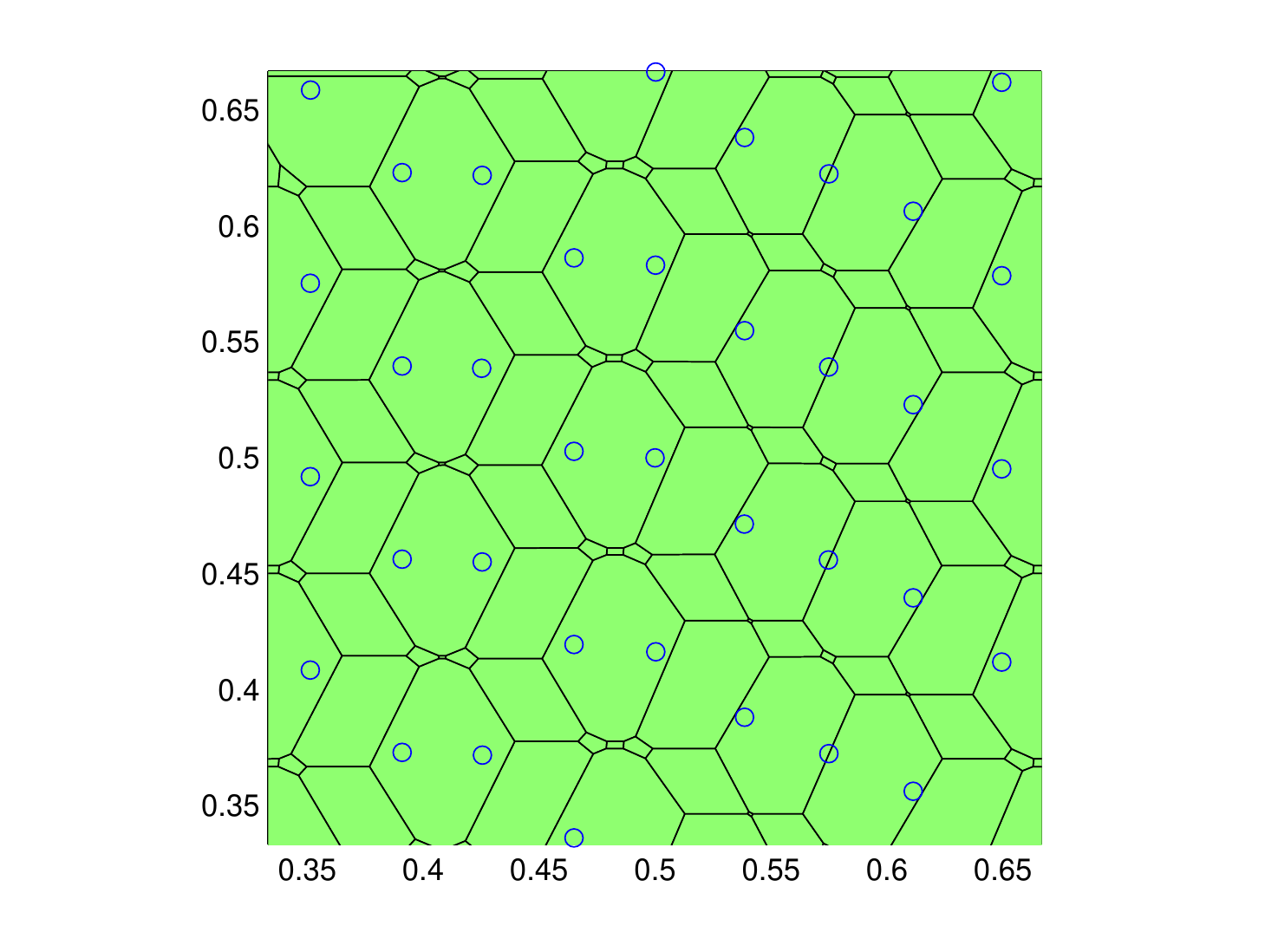}
}
\subfigure[81 agents result]{
\includegraphics[width=0.46\columnwidth]{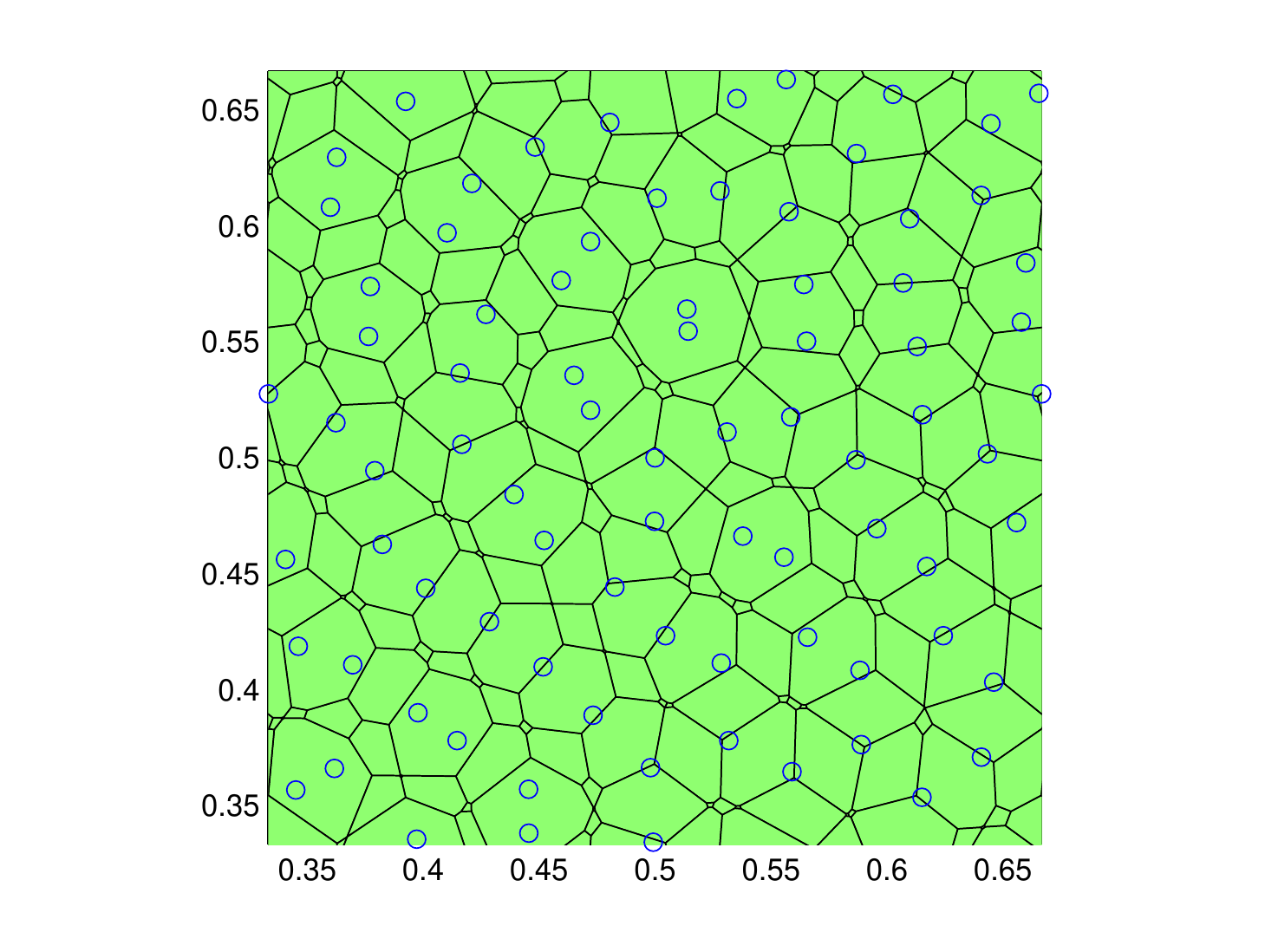}
}
\subfigure[144 agents result]{
\includegraphics[width=0.46\columnwidth]{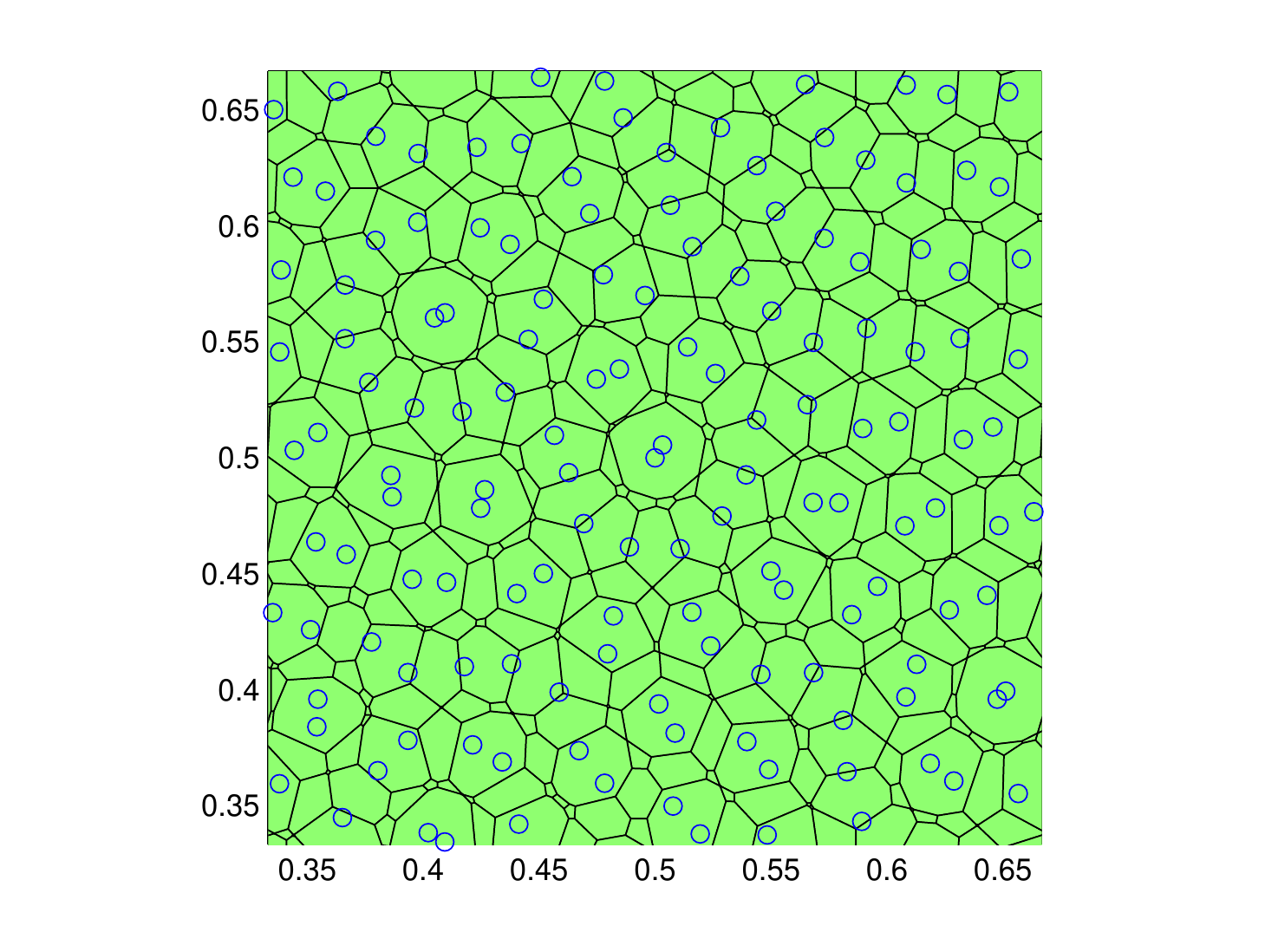}
}
\caption{A simulation result with 50 mobile sensors}\label{Unit Torus simulations}
\end{figure}

The above simulation result shows a counter-intuitive phenomenon: as the number of agents in one unit torus increases, the pattern becomes less and less regular. This is because there are several stable patterns and when those patterns combine together, the shape becomes less regular.

\section{Real world applications}   \label{Sec:application}
\label{Real world applications}

\subsection{Supermarket problem}
\label{supermarket problem definition}

As an extension of the well-known postoffice problem \cite{clarkson1985probabilistic}, the supermarket problem brings in the idea of more than one supermarket serving one cell which enables competition. When selecting sites for supermarkets, it may be considered important to promote competition among supermarkets \cite{martin2009review}.
Although supermarkets are not mobile, the optimization technique is still applicable.
Denote the set of positions of supermarkets as $(p_1, p_2,\cdots,p_n)$ in a region $Q$. For each person at position $q\in Q$, the $f(\cdot,\cdot)$ function is
$$f(\cdot,\cdot)= \max(\|q,p_i\|,\|q,p_j\|)$$
By interpreting $\phi(q)$ as  the population density at each position $q$, one can formulate the following coverage performance function
$$\int_Q \min_{(i,j)\in C} f(\cdot,\cdot) \phi(q) dq$$
where the definition of $C$ is the same as in Section \ref{order 2 section}.
To optimize the supermarket locations, one should minimize the above performance function.

\vspace{-2 mm}

\subsection{TDOA sensors and bearing-only sensors}

\vspace{-2 mm}

The above \textit{supermarket problem} is not strictly a mobile agents problem. Now we are going to consider some mobile sensing problems using the same distance function with the above.
If one uses bearing only sensors to localize targets, at least two sensors are required to obtain a measurement. Furthermore, for complete coverage, one actually needs three non-collinear bearing-only sensors to avoid the case when the sensors and target are almost collinear.
Similar to bearing-only sensors, TDOA (Time Difference of Arrival) sensors are passive sensors and at least two sensors are required to obtain a branch of a hyperbola on which a target lies. Therefore, to localize a target using TDOA sensors, three sensors at different positions are required. In this case, an order three Voronoi partition may be used to solve the problem.
\footnote{Similar to the bearing-only sensors case, for complete coverage, one actually needs four TDOA sensors to avoid undesired geometry, in which case the use of even higher order Voronoi partition is required. }

In the above example, sensors all have limited sensing range and their performance will degrade as the distance from the target to the sensor increases. If one wants to use sensor arrays to cover a large area, an order 3 coverage control problem is an appropriate tool. When detecting a target, all sensors in a cell need to receive signals from the target reliably, with the performance of monitoring a cell mainly dependent on the sensor with furthest distance to the target.  Thus the performance of a group of sensors $p_1, p_2,\cdots,p_n$ monitoring a target at $q$ can be expressed as
$$f(\cdot,\cdot,\cdot)=\max(\|q,p_i\|,\|q,p_j\|,\|q,p_k\|)$$

Suppose there is a target in a two-dimensional region $Q$. Suppose further that the probability of the target appearing at position $q\in Q$ is $\phi(q)$. Then the expected value of this localization performance measure is
$$\int_Q \min_{(i,j,k)\in C_3} f(\cdot,\cdot,\cdot) \phi(q) dq$$
where $C_3=\{i,j,k|~i,j,k\in\{1,\cdots,n\},i < j<k\}$.
Minimizing the above performance function  will give the optimal positions of sensors.

\vspace{-2 mm}

\subsection{Bi-static radar}

\vspace{-2 mm}

One possible application of our order 2 coverage control strategy involving a general $f(\cdot,\cdot)$ function is the coverage design using bi-static radars \cite{nezlin2007bistatic}. When deploying a bi-static radar, it is required that the transmitter and receiver are at different locations. We suppose that each position $p_i$ has both a transmitter and a receiver equipped but it can only receive signals from another position instead of itself. In this case, the probability of detection in one radar pulse is expressed by
\begin{equation}
\label{}
P=\int_{v_t}^\infty \frac{1}{2}I_0(\frac{\delta}{a}\sqrt{\theta})exp(-\frac{\theta-a^2/\delta^2}{2})d\theta
\end{equation}
where the term being integrated is just the probability density of the received signal strength using a common radar receiver model; see \cite{mahafza2013radar}. In particular, $v_t$ is a chosen detection threshold, $a$ is the signal amplitude, $\delta$ is the noise amplitude and $I_0$ is the modified Bessel function of the first kind and zeroth order.
The choice of $v_t$ depends on the acceptable probability of false alarm denoted by $P_{fa}$, i.e. the acceptable probability that the signal magnitude exceeds the threshold value when noise alone is present \cite{mahafza2013radar}.

For a typical bi-static radar, $a^2/\delta^2$ is the signal-to-noise ratio, which is proportional to $K/(R_1^2 R_2^2)$ where $R_1$ is the distance between the transmitter and the target, $R_2$ is the distance between the receiver and the target and $K$ depends on the radar power or antenna gain. Note $R_1$ and $R_2$ are equivalent to $\|q,p_i\|$ and $\|q,p_j\|$ in Section \ref{order 2 section}. It can be shown that
$$\frac{\partial}{\partial R_1}P \leq 0,~\frac{\partial}{\partial R_2}P \leq 0\,\,\,\,\text{and}\,\,\,\, P(R_1,R_2)=P(R_2,R_1)$$

\vspace{-1.5 mm}

Suppose there are a group of agents to be deployed in a convex area $Q$;  each agent with position $p_i$ has both transmitter and receiver on board but it can only receive signal from another agent instead of itself.
There is a target in the area and the probability of the target appearing at position $q\in Q$ is $\phi(q)$. Note there holds $\int_Q \phi(q)dq=1$. Now the probability of detection of the target at $q$ is $\max_{(i,j)\in C} P(\|q,p_i\|,\|q,p_j\|)$ and the expected value of this probability of detection is
$$E=\int_Q \max_{(i,j)\in C} P(\|q,p_i\|,\|q,p_j\|) \phi(q) dq$$
Now let $f(\cdot,\cdot)=-P(\cdot,\cdot)$, then the performance function $\mathcal{H}$ in Section \ref{order 2 section} is stated as
$$\mathcal{H}=-E=\int_Q \min_{(i,j)\in C} -P(\|q,p_i\|,\|q,p_j\|) \phi(q) dq$$

The above model is only relevant to the first detection of targets, as opposed to their localization. To achieve localization, one must build on the fact that a pair of bi-static radars only gives an ellipse for the possible positions of a target. If one needs to obtain an exact position of the target, even higher order coverage control strategy is required.

\section{Conclusions}  \label{Sec:conclusions}
In this paper, we considered a class of generalized Voronoi coverage control problems by introducing the higher order Voronoi partition concept in the coverage performance functions. This coverage problem is motivated by many real life applications which require more than one sensor to cooperate in monitoring one single cell. We focused on the order 2 Voronoi based coverage problem, and provided detailed analysis on the performance function, the controller design and controller performance (including convergence properties), supported by several simulations. The results were extended to study higher degree Lloyd maps, their connection with gradient descent algorithms, and convergence properties. In addition we considered problems of minimizing the sensor radius required to secure higher order coverage, and generalizations to the high order case of what is often termed $k$-means algorithms. Finally, we provided a number of real world scenarios where our framework can be applied.

%\begin{appendix}
%Here we  recall some basic facts in the problem of differentiation under the integral sign. Suppose $F(x,t)$, the integrand, is a function of both $x$ and $t$; suppose further $D(t)$, the domain of integration, is a function of $t$. Now the derivative of the integral with respect to $t$ at the point $t=t_0$ is
%\begin{equation}\label{sperate boundary}
%\begin{split}
%\frac{d}{dt}\int_{D(t)} F(x,t)dx \Big|_{t=t_0} =
%&\int_{D(t_0)} \frac{d}{dt} F(x,t) \Big|_{t=t_0} dx\\
%& + \frac{d}{dt}\int_{D(t)}  F(x,t_0)dx \Big|_{t=t_0}\\
%\end{split}
%\end{equation}
%where the second term is the boundary variation. Note the independent variable $x$ could be a vector instead of a scalar in this case. According to \cite{flanders1973differentiation}, the boundary variation equals
%$$\int_{\partial D(t)} F(x)v_x\cdot n_x dx$$
%where $\partial D(t)$ denotes the boundary of $D(t)$, $v_x$ denotes the direction of moving of a point $x$ on $\partial D(t)$, and $n_x$ denotes the outward unit normal vector of the point $x$ on $\partial D(t)$.
%\end{appendix}

\bibliographystyle{IEEEtran}
\bibliography{Coverage}

\end{document}